\documentclass[preprint,3p,sort&compress]{elsarticle}

\makeatletter
\def\ps@pprintTitle{%
 \let\@oddhead\@empty
 \let\@evenhead\@empty
 \def\@oddfoot{}%
 \let\@evenfoot\@oddfoot}
\makeatother
\usepackage{amssymb}
\usepackage{amsmath}
\usepackage{xspace}

\usepackage[hidelinks]{hyperref} %Remove the ugly boxes
\hypersetup{colorlinks,urlcolor=blue,linkcolor=blue,citecolor=blue,filecolor=blue}
\newcommand{\NORSE}{\texttt{NORSE}\xspace}
\newcommand{\CODE}{\texttt{CODE}\xspace}

\newcommand{\sub}[1]{_{\mathrm{#1}}}
\renewcommand{\d}{\mbox{d}}
\newcommand{\dthreep}{\d^3 p}
\newcommand{\dep}{\d p}

\newcommand{\ep}{\mathbf{e}_{p}}
\newcommand{\exi}{\mathbf{e}_{\xi}}

\newcommand{\prefactor}{C}
\newcommand{\Wbar}{\overline{W}}
\newcommand{\Umin}{\Upsilon_{-}}
\newcommand{\Uplus}{\Upsilon_{+}}
\newcommand{\Uzero}{\Upsilon_0}
\newcommand{\Uone}{\Upsilon_1}
\newcommand{\Utwo}{\Upsilon_2}
\newcommand{\Pzero}{\Pi_0}
\newcommand{\Pone}{\Pi_1}
\newcommand{\zeff}{Z\sub{eff}}
\newcommand{\epso}{\epsilon_{0}}
\newcommand{\lnL}{\ln\Lambda}
\newcommand{\me}{m\sub{e}}
\newcommand{\pder}[2]{\ensuremath{\frac{\partial #1}{\partial #2}}}
\newcommand{\pdersec}[2]{\ensuremath{\frac{\partial^2 #1}{\partial {#2}^2}}}
\newcommand{\power}[1]{\!\cdot\! 10^{#1}}
\newcommand{\Eq}[1]{Eq.~\eqref{#1}}
\newcommand{\Fig}[1]{Fig.~\ref{#1}}
\newcommand{\Ref}[1]{Ref.~\cite{#1}}

\renewcommand{\mod}[1]{#1}%{\textcolor{cyan}{#1}}%{#1}

\begin{document}

\begin{frontmatter}

\title{NORSE: A solver for the relativistic non-linear Fokker-Planck equation for electrons in a homogeneous plasma}

\author[Chalmers]{A. Stahl\corref{CorrAuthor}}
\author[Maryland]{M. Landreman}
\author[Chalmers]{O. Embr\'eus}
\author[Chalmers]{T. F\"ul\"op}

\address[Chalmers]{Chalmers University of Technology, G\"oteborg,  Sweden}
\address[Maryland]{University of Maryland, College Park, MD, USA}
\cortext[CorrAuthor]{stahla@chalmers.se}

%%%%%%%%%%%%%%%%%%%%%%%%%%%%%%%%%%%%%%%%%%%%%%%%%%%%%%%%%%%%%%%%%%%%%%%%%%%%

\begin{abstract}
Energetic electrons are of interest in many types of plasmas, however previous modeling of their properties has been restricted to the use of linear Fokker-Planck collision operators or non-relativistic formulations.  Here, we describe a fully non-linear kinetic-equation solver, capable of handling large electric-field strengths (compared to the Dreicer field) and relativistic temperatures. This tool allows modeling of the momentum-space dynamics of the electrons in cases where strong departures from Maxwellian distributions may arise. As an example, we consider electron runaway in magnetic-confinement fusion plasmas and describe a transition to electron slide-away at field strengths significantly lower than previously predicted.
\end{abstract}

\begin{keyword}
Non-linear relativistic Fokker-Planck equation \sep kinetic plasma theory \sep energetic electrons \sep runaway electrons

%% PACS codes here, in the form: 
\PACS 52.25.Dg \sep 52.65.Ff

%% MSC codes here, in the form: \MSC code \sep code
%% or \MSC[2008] code \sep code (2000 is the default)
\end{keyword}

\end{frontmatter}

%\linenumbers

%%%%%%%%%%%%%%%%%%%%%%%%%%%%%%%%%%%%%%%%%%%%%%%%%%%%%%%%%%%%%%%%%%%%%%%%%%%%
%%%%%%%%%%%%%%%%%%%%%%%%%%%%%%%%%%%%%%%%%%%%%%%%%%%%%%%%%%%%%%%%%%%%%%%%%%%%

% Computer program descriptions should contain the following
% PROGRAM SUMMARY.

{\bf PROGRAM SUMMARY}

\begin{small}
\noindent
{\em Program title: NORSE}                                          \\
{\em Program Files doi:} \href{http://dx.doi.org/10.17632/86wmgj758w.1}{http://dx.doi.org/10.17632/86wmgj758w.1}									\\
{\em Licensing provisions: GPLv3}                                   \\
{\em Programming language: Matlab}                                   \\
{\em Nature of problem:}\\
  Solves the Fokker-Planck equation for electrons in 2D momentum space in a homogeneous plasma (allowing for magnetization), using a relativistic non-linear electron-electron collision operator. Electric-field acceleration, synchrotron-radiation-reaction losses, as well as heat and particle sources are included. Scenarios with time-dependent plasma parameters can be studied.\\
{\em Solution method:}\\
  The kinetic equation is represented on a non-uniform 2D finite-difference grid and is evolved using a linearly implicit time-advancement scheme. A mixed finite-difference--Legendre-mode representation is used to obtain the relativistic potentials (analogous to the non-relativistic Rosenbluth potentials) from the distribution.\\
   \\

\end{small}

%%%%%%%%%%%%%%%%%%%%%%%%%%%%%%%%%%%%%%%%%%%%%%%%%%%%%%%%%%%%%%%%%%%%%%%%%%%%
%%%%%%%%%%%%%%%%%%%%%%%%%%%%%%%%%%%%%%%%%%%%%%%%%%%%%%%%%%%%%%%%%%%%%%%%%%%%
%%%%%%%%%%%%%%%%%%%%%%%%%%%%%%%%%%%%%%%%%%%%%%%%%%%%%%%%%%%%%%%%%%%%%%%%%%%%

\section{Introduction}
Energetic electrons, having speeds significantly larger than the
average speed of the thermal population, are ubiquitous in
plasmas. Examples are found for instance in the solar corona
\cite{Roberts1998} and wind \cite{Pilipp1987a}, and in solar flares
\cite{VanAllen1965,Holman1985}; in the ionosphere of the Earth \cite{Bell1995} and
lightning discharges \cite{Gurevich1994}; as well as in laboratory
laser-plasma accelerators \cite{Malka2008} and inertial
\cite{Tabak2005} and magnetic-confinement \cite{Helander2002} fusion
plasmas. In the latter case, understanding the dynamics of the
energetic electrons is of particular concern, as so-called
\emph{runaway electrons} \cite{Dreicer1959,Dreicer1960} generated
during \emph{disruptions} -- events where the plasma rapidly cools and
strong electric fields are induced -- have the potential to cause
severe damage to a tokamak fusion reactor. This problem is only
expected to become more severe in future devices since the runaway generation is exponentially sensitive to the available plasma current
\cite{RosenbluthPutvinski1997,Boozer2015}.

In a spatially homogeneous plasma, the main processes influencing
energetic-electron dynamics are: the presence of an accelerating
electric field; magnetization (causing directed motion); Coulomb collisions; dynamic changes in plasma parameters such
as the temperature; radiative losses (associated with synchrotron and fast-electron bremsstrahlung emission); and wave-particle interaction. The combined influence of these processes has been shown to lead to phenomena such as bump-on-tail formation \cite{HirvijokiBump2015,DeckerBump2016} and local isotropization \cite{Pokol2008,Pokol2014} in the high-energy
tail of strongly anisotropic electron populations. Since analytic
treatment is possible only in special cases, the evolution of the
electron distribution function $f$ must in general be studied using
kinetic simulations.

Many numerical tools solve the kinetic equation for $f$, taking
some subset of the processes mentioned above into account. In
collisional plasmas, the Fokker-Planck operator describing the Coulomb
collisions is the main source of complexity in the problem, which in
general is described by a stiff integro-differential diffusion
equation, and numerical treatments can be broadly categorized based on
the level of sophistication of the collision operator employed.  A
number of continuum tools have been developed that use linearized
(around a Maxwellian) collision operators, especially in the case of fully
non-relativistic problems, but also in scenarios where the electrons are allowed to reach relativistic energies \cite{Chiu1998,Decker2004,CODEPaper2014}. In addition,
several fully non-linear tools are available for non-relativistic
scenarios
\cite{McCoy1981,Chacon2000,Pareschi2000,Weng2006,Pataki2011,Yoon2014,Taitano2015,HirvijokiRBF2015};
however, to our knowledge, no tool treats the relativistic non-linear collision operator in its entirety. Both the integrated-tokamak-modeling tool \texttt{TASK} and \texttt{CQL3D} (which is focused on heating and current drive in tokamaks) successfully implement the first few Legendre modes of the relativistic non-linear collision operator
\cite{Nuga2011,Petrov2009}. While this approach guarantees the conservation of density, momentum and energy, it cannot resolve fine structures in the momentum-space distribution, making it unsuitable for accurate study of the fast-electron dynamics. In \texttt{CQL3D}, the implementation is general and therefore in principle supports the use of any number of modes, however in practice, the maximum number of modes cannot exceed 3 to 5 because of numerical problems \cite{Petrov2009}.

In the magnetic-fusion community in particular, there is a pressing
need for a tool with the ability to handle situations where relativistic particles comprise a significant part of the overall electron distribution, as these are the situations of greatest danger to the integrity of the
fusion device \cite{Boozer2015}. Such scenarios arise primarily when
the electric field magnitude is (at least) a significant fraction of the so-called \emph{Dreicer field} \cite{Dreicer1959},
$E\sub{D}=ne^3\lnL/4\pi\epso^2T$; where $n$, $T$ and $-e$ are the
electron number density, temperature and charge, $\lnL$ is the Coulomb
logarithm, and $\epso$ is the vacuum permittivity. For such
field strengths, the electric field overcomes the maximum collisional friction force affecting the electrons. However, also for $E<E\sub{D}$, the distortion of the distribution can become substantial, leading to the break-down of linearized codes. In contrast, the so-called \emph{critical field} $E\sub{c}=\Theta E\sub{D}$, with
$\Theta=T/\me c^2$ the bulk temperature normalized to the electron rest
mass, is equivalent to the minimum collisional friction force
experienced by highly relativistic electrons. It therefore describes
the weakest field at which runaway-electron generation can occur
\cite{ConnorHastie1975}, since the accelerating force must overcome the friction (and the latter decreases with increasing particle energy). For non-relativistic bulk-electron
temperatures, $E\sub{c}$ is much less than $E\sub{D}$ and runaway-electron generation
can in general be studied using a linearized treatment since the electric field can fulfill both $E>E\sub{c}$ and $E\ll E\sub{D}$ simultaneously. A fully
non-linear relativistic tool is however needed in the scenarios of
highest importance, where the runaway population becomes comparable to the thermal population or the electric field is of order $E\sub{D}$.

In this paper, we describe such a tool: the new finite-difference code
\NORSE (NOn-linear Relativistic Solver for Electrons), which
efficiently solves the kinetic equation in 2D momentum space. \NORSE
includes a fully relativistic non-linear Fokker-Planck operator for electron-electron collisions \cite{BeliaevBudker1956,BraamsKarney1987,BraamsKarney1989} 
and synchrotron-radiation-reaction effects \cite{Andersson2001,HirvijokiBump2015}. Time-dependent plasma parameters make the investigation
of dynamic scenarios possible.

With a non-linear treatment, the distribution is not restricted to
being approximately Maxwellian, and strong electric fields (compared to $E\sub{D}$) can therefore be applied. However, if the distribution departs strongly from a Maxwellian, concepts such as temperature and the thermal collision time are not well defined. In many scenarios of practical interest, the distribution will nevertheless stay close to collisional equilibrium, and we will make use of familiar concepts where appropriate.

The kinetic equation and the operators for the various mechanisms
mentioned above are discussed in Section \ref{sec:Kinetic_eq}. The
numerical implementation is then outlined in Section
\ref{sec:Numerical_method} and validated in Section \ref{sec:Tests}
through a comparison to previous work in several limits. Finally, in
Section \ref{sec:Application--runaway} we use \NORSE to investigate
the properties of strongly distorted electron distributions and the condition for electron slide-away.

%%%%%%%%%%%%%%%%%%%%%%%%%%%%%%%%%%%%%%%%%%%%%%%%%%%%%%%%%%%%%%%%%%%%%%%%%%%%
%%%%%%%%%%%%%%%%%%%%%%%%%%%%%%%%%%%%%%%%%%%%%%%%%%%%%%%%%%%%%%%%%%%%%%%%%%%%
%%%%%%%%%%%%%%%%%%%%%%%%%%%%%%%%%%%%%%%%%%%%%%%%%%%%%%%%%%%%%%%%%%%%%%%%%%%%

\section{Kinetic equation}
\label{sec:Kinetic_eq}
To study the momentum-space dynamics of energetic electrons, we will
solve the kinetic equation neglecting any spatial dependence. A point
in the 3D momentum space is represented in spherical coordinates by
$\mathbf{p} = (p,\xi,\varphi)$, with $p=\gamma v/c$ the magnitude of
the (normalized) momentum, $\xi=p_{\|}/p$ the cosine of the polar
angle, and $\varphi$ the azimuthal angle. Here $v$ is the speed of the
particle, $c$ is the speed of light, and $\gamma$ is the relativistic
mass factor. The spherical symmetry of our problem is broken by the
presence of the electric field $\mathbf{E}$, and we therefore let the
electric field define the parallel direction. If the plasma is magnetized, only the electric-field component parallel to $\mathbf{B}$ contributes to the acceleration (i.e. $\mathbf{B} \| \mathbf{E}$), in which case $\xi$ is the cosine of the pitch--angle and $\varphi$ is the gyro angle. We will assume the
electron distribution function $f=f(t; p,\xi)$ to be independent of
$\varphi$, reducing the problem to a two-dimensional one. 

The kinetic equation describing the evolution of
$f$ can be written as
\begin{equation}
	\pder{f}{t} 
	-\frac{e\mathbf{E}}{\me c} 
		\cdot \pder{f}{\mathbf{p}} 
	+\pder{}{\mathbf{p}}\cdot
		\left(
			\mathbf{F}\sub{s}f
		\right)
	=C\{f\} + S,
	\label{eq:kinetic_eq}
\end{equation}
where $\mathbf{F}\sub{s}$ is the synchrotron-radiation-reaction force (in the presence of a magnetic field), $C\{f\}$ is the Fokker-Planck collision operator describing microscopic Coulomb interactions between the plasma particles, and $S$ denotes sources and sinks (of for instance heat or particles). The distribution $f$ satisfies $n=\int\dthreep\,f$, with $n$ the number density of electrons. 
 
The parallel component of the momentum-space gradient, appearing in the term describing the Hamiltonian motion of the electrons due to the electric field, becomes 
\begin{equation}
	\frac{\mathbf{E}}{E} \cdot \pder{f}{\mathbf{p}} 
	= \left(
		\xi\pder{f}{p} + \frac{1-\xi^{2}}{p}\pder{f}{\xi}
	  \right).
\end{equation}
In what follows, we will detail the synchrotron-radiation-reaction and collision terms of \Eq{eq:kinetic_eq}, as well as the various source terms.

%%%%%%%%%%%%%%%%%%%%%%%%%%%%%%%%%%%%%%%%%%%%%%%%%%%%%%%%%%

\subsection{Synchrotron-radiation reaction}
The reaction force experienced by electrons emitting synchrotron radiation can be derived from the Lorentz--Abraham--Dirac force. In a homogeneous plasma, it can be written as (see for instance \cite{HirvijokiBump2015} and references therein)
\begin{align}
\pder{}{\mathbf{p}} \cdot \left( \mathbf{F}_{S}f \right)
&=-\frac{1}{p^{2}}\pder{}p
		\left(
			\frac{\gamma p^{3}(1-\xi^{2})}{\tau_{r}}f
		\right)
+\pder{}{\xi}
		\left(
			\frac{\xi(1-\xi^{2})}{\gamma\tau_{r}}f
		\right)\notag\\
&=-\frac{1}{\tau_{r}}\frac{1-\xi^{2}}{\gamma}
	\left[
		\gamma^{2}p\pder fp-\xi\pder f{\xi}
		+\left(
			4p^{2}+\frac{2}{1-\xi^{2}}
		\right)f
	\right],
	\label{eq:synch_force_components}
\end{align}
where
\begin{equation}
\tau_{r}=\frac{6\pi\epso(\me c)^{3}}{e^{4}B^{2}}
\end{equation}
is the radiation time-scale. Here $B$ is the magnetic field strength. 

The total synchrotron power emitted by a relativistic particle is proportional to $p_{\perp}^2=p^2(1-\xi^2)$, and the back-reaction experienced by the electrons therefore increases with perpendicular momentum. The efficacy of the synchrotron-radiation reaction is thus closely linked to collisional pitch--angle scattering, which can redistribute parallel momentum gained from the accelerating field.

%%%%%%%%%%%%%%%%%%%%%%%%%%%%%%%%%%%%%%%%%%%%%%%%%%%%%%%%%%%

\subsection{Electron-ion collision operator}
In a fully ionized plasma, the collision operator $C$ contains contributions from collisions with electrons ($C\sub{ee}$) and ions ($C\sub{ei}$):
\begin{equation}
	C\{f\} = C\sub{ee}\{f\} + C\sub{ei}\{f\}.
\end{equation}
The electron-electron collision operator is the main source of complexity in our problem, and will be discussed in Section \ref{sec:ee_coll_op}. In contrast, we will assume a stationary, Maxwellian ion population. This, together with the mass difference between the species involved in the collision, significantly simplifies the operator for electron-ion collisions (unless the ratio between ion and electron temperatures is comparable to their mass ratio). In the ion rest-frame, the operator is \citep{HelanderSigmar2002}: 
\begin{equation}
C\sub{ei}\{f\}
\simeq \zeff\frac{\nu\gamma}{p^{3}}\mathcal{L}\{f\}
=\zeff\frac{\nu\gamma}{p^{3}}
	\left[
		\frac{1}{2}\pder{}{\xi}(1-\xi^2)\pder{f}{\xi}		
	\right],
\end{equation}
where $\zeff=n^{-1}\sum_j n_j Z_j^2$ is the effective charge (with the sum taken over all ion species $j$), $\mathcal{L}$ is the Lorentz scattering operator, and
\begin{equation}
\nu=\frac{ne^{4}\lnL}{4\pi\epso^{2}m_{e}^{2}c^{3}}
\end{equation}
is the collision frequency for relativistic electrons. The operator $C\sub{ei}$ describes pitch--angle scattering, but no change to the magnitude of the electron momentum. This is because the ions are assumed to be much heavier than the electrons (i.e. $m\sub{i} \gg \gamma \me$), so that the energy lost by the electrons through collisions can be neglected.

%%%%%%%%%%%%%%%%%%%%%%%%%%%%%%%%%%%%%%%%%%%%%%%%%%%%%%%%%%%

\subsection{Electron-electron collision operator}
\label{sec:ee_coll_op}

To describe electron-electron collisions, we will use the fully relativistic non-linear collision operator of Beliaev \& Budker \citep{BeliaevBudker1956}, in the form developed
by Braams \& Karney \citep{BraamsKarney1987,BraamsKarney1989}. The operator is valid for collisions between arbitrary species of arbitrary energy (i.e. the bulk population is not required to be non-relativistic). For electron-electron collisions, it takes the form \citep{BraamsKarney1989} 
\begin{equation}
C\sub{ee}\{f\}
=\alpha \pder{}{\mathbf{p}}\cdot
	\left(
		\mathbb{D}\cdot\pder f{\mathbf{p}}-\mathbf{F}f
	\right)
\label{eq:Cee_short}
\end{equation}
where $\alpha=4\pi\nu/n$, $\mathbb{D}$ is the diffusion tensor and $\mathbf{F}$ is the friction vector. These are given by
\begin{align}
\mathbb{D} &= \gamma^{-1}
	\left[
		\mathbb{L}\Umin
		- ( \mathbb{I}+\mathbf{p}\mathbf{p} ) \Uplus
	\right],\\
\mathbf{F} &= \gamma^{-1}\mathbf{K}\Pi, 
\end{align}
where $\mathbb{I}$ is the unit tensor and $\mathbb{L}$ and $\mathbf{K}$ are defined by
\begin{align}
\mathbb{L}\Umin 
&=
	\left(
		\mathbb{I}+\mathbf{p}\mathbf{p}
	\right)
		\cdot\frac{\partial^{2} \Umin}{\partial\mathbf{p}\partial\mathbf{p}}\cdot
	\left(
		\mathbb{I}+\mathbf{p}\mathbf{p}
	\right)	
  + \left(
		\mathbb{I}+\mathbf{p}\mathbf{p}
	\right)
	\left(
	\mathbf{p}\cdot\pder{\Umin}{\mathbf{p}}
	\right),\\
\mathbf{K}\Pi
&= \left(
		\mathbb{I}+\mathbf{p}\mathbf{p}
	\right)
	\cdot\pder{\Pi}{\mathbf{p}}.
\end{align}
Here, $\Umin$, $\Uplus$ and $\Pi$ are linear combinations of potential functions, given by
\begin{equation}
\Umin=4\Upsilon_{2}-\Upsilon_{1},\qquad\Upsilon_{+}=4\Upsilon_{2}+\Upsilon_{1},\qquad\Pi=2\Pi_{1}-\Pi_{0},
\end{equation}
where we denote the five potentials introduced by Braams \& Karney as $\Uzero$, $\Uone$, $\Utwo$, $\Pzero$ and $\Pone$. These are defined using the differential operator 
\begin{equation}
L_{a}\Psi  =\left(\mathbb{I}+\mathbf{p}\mathbf{p}\right):\frac{\partial^{2}\Psi}{\partial\mathbf{p}\partial\mathbf{p}}+3\mathbf{p}\cdot\pder{\Psi}{\mathbf{p}}+\left(1-a^{2}\right)\Psi,
\label{eq:L_a}
\end{equation}
through
\begin{equation}
L_{0}\Upsilon_{0} =f,\quad 
L_{2}\Upsilon_{1}=\Upsilon_{0},\quad L_{2}\Upsilon_{2}=\Upsilon_{1},\quad 
L_{1}\Pi_{0}=f,\quad 
L_{1}\Pi_{1}=\Pi_{0}.
\label{eq:potentials_relations}
\end{equation}
The five potentials are analogous to the two Rosenbluth potentials $g$ and $h$ in the non-relativistic case \cite{Rosenbluth1957}, and reduce to these in the appropriate limit. The notation (which differs from that in \Ref{BraamsKarney1989}, see \ref{app:potentials_bcs}) has been chosen to highlight the existence of two ``branches'' of potentials, distinguished by the application of different $L_a$ operators. Crucially, in the non-relativistic limit (as $L_a$ reduces to the Laplace operator), $\Uzero=\Pzero\to h$ and $\Uone=\Pone\to g$.

A sketch of the derivation of the explicit expressions obtained in our coordinate system is given in \ref{app:Cee_der}; here we list only the final result, with the terms grouped according to the derivative of $f$. The collision operator can be written as
\begin{equation}
\frac{C\sub{ee}\{f\}}{\alpha}
=\prefactor^{(p^{2})} \pdersec{f}{p}
+\prefactor^{(p)} \pder{f}{p}
+\prefactor^{(\xi^{2})} \pdersec{f}{\xi}
+\prefactor^{(\xi)} \pder{f}{\xi}
+\prefactor^{(p\xi)} \pder{^{2}f}{p\partial\xi}
+\prefactor^{(f)}\! f,
\label{eq:Cee_components}
\end{equation}
with pre-factors $\prefactor^{(i)}$ given by
\begin{align}
\prefactor^{(p^{2})} &= 
	\gamma(8\Utwo-\Uzero)
	-2\frac{\gamma^{3}}{p} \pder{\Umin}{p}
	-\frac{\gamma(1-\xi^{2})}{p^{2}} \pdersec{\Umin}{\xi}
	+2\frac{\gamma\xi}{p^{2}} \pder{\Umin}{\xi}, \label{eq:p2_term}\\
\prefactor^{(p)} &= 	
	\frac{1}{\gamma p}(2+3p^{2})(8\Upsilon_{2}-\Upsilon_{0})-16\gamma\pder{\Upsilon_{2}}p+6\gamma\pder{\Upsilon_{1}}p-\gamma\pder{\Upsilon_{0}}p-2\frac{\gamma^{3}}{p}\left(\pdersec{\Upsilon_{-}}p+\frac{1}{p}\pder{\Upsilon_{-}}p\right)\nonumber \\
 & +\frac{1}{\gamma p}\left(2+\frac{1}{p^{2}}\right)\left(2\xi\pder{\Upsilon_{-}}{\xi}-(1-\xi^{2})\pdersec{\Upsilon_{-}}{\xi}\right)-\gamma\pder{\Pi}p, \\ 
\prefactor^{(\xi^{2})} &=
	\frac{1-\xi^{2}}{\gamma p^{2}}\left(\frac{\gamma^{2}}{p}\pder{\Upsilon_{-}}p+\frac{1}{p^{2}}\left[(1-\xi^{2})\pdersec{\Upsilon_{-}}{\xi}-\xi\pder{\Upsilon_{-}}{\xi}\right]-\Upsilon_{+}\right),\\ 
\prefactor^{(\xi)} &= 
	-\frac{\xi(1-\xi^{2})}{\gamma p^{4}}\pdersec{\Upsilon_{-}}{\xi}-2\frac{\gamma(1-\xi^{2})}{p^{3}}\frac{\partial^{2}\Upsilon_{-}}{\partial p\partial\xi}-2\frac{\gamma\xi}{p^{3}}\pder{\Upsilon_{-}}p\nonumber \\
 & +\left(\frac{2}{\gamma p^{4}}+3\frac{1-\xi^{2}}{\gamma p^{2}}\right)\pder{\Upsilon_{-}}{\xi}-\frac{1-\xi^{2}}{\gamma p^{2}}\left(4\pder{\Upsilon_{2}}{\xi}-3\pder{\Upsilon_{1}}{\xi}+\pder{\Upsilon_{0}}{\xi}+\pder{\Pi}{\xi}\right)+2\frac{\xi}{\gamma p^{2}}\Upsilon_{+}, \\ 	
\prefactor^{(p\xi)} &=
	2\frac{\gamma(1-\xi^{2})}{p^{3}}\left[p\frac{\partial^{2}\Upsilon_{-}}{\partial p\partial\xi}-\pder{\Upsilon_{-}}{\xi}\right], \\ 
\prefactor^{(f)} &=
	-\gamma\pdersec{\Pi}p-\frac{1}{\gamma p}\left(2+3p^{2}\right)\pder{\Pi}p-\frac{1-\xi^{2}}{\gamma p^{2}}\pdersec{\Pi}{\xi}+2\frac{\xi}{\gamma p^{2}}\pder{\Pi}{\xi}. \label{eq:f_term}
\end{align}

%%%%%%%%%%%%%%%%%%%%%%%%%%%%%%%%%%%%%%%%%%%%%%%%%%%%%%%%%%%

\subsection{Heat and particle sources}
\label{sec:sources}
A strong electric field is a source of energy that quickly heats the
distribution function. In contrast to a linearized treatment (where this heat must be removed to ensure the validity of the linearization), this energy source is automatically accounted for in the non-linear solution. Sometimes, it is however of interest to remove the excess heat from the bulk as it is applied. In reality, the bulk temperature is not always increasing during fast-particle generation, for instance because of energy loss due to radiation emission or heat conduction. A heat sink also serves as a way to vary the temperature of the thermal population, which makes it possible to model dynamic scenarios where the plasma parameters change on a time scale similar to that of the acceleration dynamics. To be able to model density changes, a particle source must also be included.

An advantageous way to formulate a heat sink is to write it in divergence form
\begin{equation}
\pder{}{\mathbf{p}} \cdot \big( k\sub{h} \mathbf{S}\sub{h} f \big)
= k\sub{h}
	\left(
		\frac{2}{p}S\sub{h}(p)
		+\pder{S\sub{h}(p)}{p}
		+S\sub{h}(p)\pder{}{p}
	\right) f,
\end{equation}
since it will then automatically conserve particles (before discretization). Here $\mathbf{S}\sub{h}(p)$ is an isotropic function of \mod{momentum, with $S\sub{h}$ its $p$-component}. $k\sub{h}$ is the magnitude, to be determined. \mod{In practice, the exact momentum-space shape of the sink depends on the processes responsible for the heat loss. A detailed investigation of this is left for future work; here we let $S\sub{h}$ have the shape of a Maxwellian for simplicity.} 

Apart from the electric-field term, synchrotron-radiation reaction also \mod{changes the total heat content of the distribution by removing energy}, primarily at large \mod{particle} momenta. \mod{However, the momentum-space region $\boldsymbol{\Omega}$ of interest need not necessarily encompass the entire computational domain. For instance, it is sometimes desirable to maintain a fixed energy content in the thermal population, while simultaneously allowing the energetic particles to gain energy. Physically, this corresponds to heat sinks that only affect slow particles. In such cases, collisions may also transfer energy into or out of $\boldsymbol{\Omega}$.} The total energy change \mod{in $\boldsymbol{\Omega}$} can thus be written as
\begin{equation}
\frac{\d W}{\d t} 
= \me c^2 \int_{\boldsymbol{\Omega}}\dthreep\, (\gamma-1)\!
	\left(
		-\frac{e\mathbf{E}}{\me c} 
		\cdot \pder{f}{\mathbf{p}} 
		+\pder{}{\mathbf{p}}\cdot
			\left(
				\mathbf{F}\sub{s}f
			\right)
		\mod{-C\{f\}}
		+k\sub{h}\pder{}{\mathbf{p}}\cdot(\mathbf{S}\sub{h}f)
	\right),
\end{equation}
from which $k\sub{h}$ can be determined in each time step by demanding that $\d W/\d t=0$ \mod{(for an ideal heat sink)}. (Note that this approach does not automatically enforce energy conservation after discretization; only physical sources of heat are taken into account.  If desirable, numerical heating caused by the discretization can be eliminated by instead requiring the numerically calculated energy moment of $f$ to be constant.\mod{)} 

The same heat source can be used to induce changes to the bulk temperature, but in this case the magnitude $k\sub{h}$ is calculated differently. We note that the relativistic equilibrium Maxwell-J\"uttner distribution
\begin{equation}
f\sub{M}(p) = \frac{n}{4\pi\Theta K_{2}\bigl(1/\Theta\bigr)}
\exp\left( -\frac{\gamma(p)}{\Theta} \right),
\end{equation}
where $K_{\nu}(x)$ is the modified Bessel function of the second kind (and order $\nu$), has the energy moment
\begin{equation}
W(\Theta) = \frac{\me c^{2} n}{\Theta K_2(1/\Theta)}
\int_{0}^{p_{\max,\boldsymbol{\Omega}}}\!\!\!\d p\,p^{2} (\gamma-1)\exp
\left(-\frac{\gamma}{\Theta}\right). 
\end{equation}
The magnitude $k\sub{h}$ can be determined from the requirement that the energy supplied by the heat sink should equal $W(\Theta_2)-W(\Theta_1)$ for two temperatures $\Theta_1$ and $\Theta_2$ at subsequent time steps. Here $p_{\max,\boldsymbol{\Omega}}$ denotes the upper boundary of $\boldsymbol{\Omega}$ in $p$; if $p_{\max,\boldsymbol{\Omega}}\to \infty$, the above integral can be evaluated analytically, yielding
\begin{equation}
W(\Theta) = \me c^{2} n 
	\left(
		\frac{K_3(1/\Theta)}{K_2(1/\Theta)} -1 -\Theta
 	\right)
\equiv \me c^2 n \Wbar (\Theta). 
\label{eq:W_Maxwellian}
\end{equation}

Changes to the density can be introduced using a particle source of the form
\begin{equation}
	S\sub{p} = k\sub{p} \left[\frac{\gamma-1}{\Theta} + a\sub{p}(\Theta) \right]f\sub{M}; 
	\label{eq:particle_sink}
\end{equation}
a linear combination of the energy and density moments of a Maxwellian (with an overall scaling factor $k\sub{p}$ analogous to $k\sub{h}$). The quantity $a\sub{p}$ can be determined from the constraint that the energy moment of $S\sub{p}$ should vanish (so that the source supplies particles, but no heat), giving
\begin{equation}
a\sub{p}(\Theta) = - \frac{1}{\Theta} 
	\frac{
		\int_{0}^{\infty}\!\!\!\d p\,p^{2} (\gamma-1)^2 \exp(-\gamma/\Theta)
		}{
		\int_{0}^{\infty}\!\!\!\d p\,p^{2} (\gamma-1) \exp(-\gamma/\Theta)
		}
= \frac{2}{\Theta} - \frac{3(1+\Theta)}{\Wbar(\Theta)} - 3.
\end{equation} 
As $\Theta\to 0$, the non-relativistic limit $a\sub{p}=-5/2$ is recovered, whereas in the  ultra-relativistic case ($\Theta\ggg 1$), $a\sub{p}\to -4$. The density moment $n\sub{p}$ of the source is given by
\begin{equation}
\frac{n\sub{p}}{n} 
= k\sub{p} 
	\left[
		\frac{\Wbar(\Theta)}{\Theta} 
		+ a(p)
	\right]
= k\sub{p}
	\left[
		\frac{\Wbar(\Theta)+2}{\Theta} 
		- 3\frac{\Wbar(\Theta) + 1+\Theta}{\Wbar(\Theta)}
	\right],
\end{equation}
from which the magnitude $k\sub{p}$ that gives a desired density change can be determined. The bracket takes the asymptotic value -1 at both $\Theta \to 0$ and $\Theta \ggg 1$, but reaches a minimum of $-1.18$ for intermediate temperatures.

%%%%%%%%%%%%%%%%%%%%%%%%%%%%%%%%%%%%%%%%%%%%%%%%%%%%%%%%%%%%%%%%%%%%%%%%%%%%
%%%%%%%%%%%%%%%%%%%%%%%%%%%%%%%%%%%%%%%%%%%%%%%%%%%%%%%%%%%%%%%%%%%%%%%%%%%%
%%%%%%%%%%%%%%%%%%%%%%%%%%%%%%%%%%%%%%%%%%%%%%%%%%%%%%%%%%%%%%%%%%%%%%%%%%%%

\section{Numerical method}
\label{sec:Numerical_method}

\subsection{Discretization}

We choose to represent the distribution $f$ on a two-dimensional finite-difference grid in $p$ and $\xi$, and use a 5-point stencil to discretize the momentum-space derivatives. Moments of the distribution and other integrals are calculated using a composite Simpson's rule. The grid points can be chosen non-uniformly in both $p$ and $\xi$, making it possible to efficiently resolve both a Maxwellian bulk (assuming there is one) and a high-energy tail. Specifically, the $p$ grid should preferably be densely spaced for small $p$ to resolve the bulk, but since the tail generally varies over larger momentum scales, coarser spacing can be used at larger momenta to reduce the computational expense. Similarly, the $\xi$ grid should be densely spaced close to $\xi=1$ (the parallel direction) to resolve the tail drawn out by the electric field. Alternatively, in scenarios without a preferred direction of acceleration, a grid which gives a uniform spacing in the polar angle ($\arccos\xi$) is often appropriate. Due to the polar nature of the coordinate system, the point at $p=0$ is special; the value of the distribution at $p=0$ should be independent of $\xi$. The total number of grid points is thus $N_{\xi}\times(N_p-1)+1$, with $N_p$ and $N_{\xi}$ the number of grid points in the respective coordinate, and a single (rather than $N_{\xi}$) grid point appropriately describes the system at $p=0$.

For the calculation of the potentials $\Upsilon_{i}$ and $\Pi_{i}$ (here collectively denoted by $\Psi$), it is advantageous to decompose the $\xi$ coordinate in Legendre modes (rather than use a finite difference grid), since these are eigenfunctions of the collision operator. The distribution and potentials are then written as
\begin{equation}
f(p,\xi) = \sum_{l=0}^{N_{l}} f_l(p)P_l(\xi), \qquad
\Psi(p,\xi) = \sum_{l=0}^{N_{l}} \Psi_l(p)P_l(\xi), 
\label{eq:f_leg_modes}
\end{equation} 
where $P_l$ is the $l$th Legendre polynomial. The potentials are integral moments of the distribution function, and their calculation is a smoothing operation. Therefore, a small number $N_{l}$ of Legendre modes typically suffices to accurately describe the potentials, unless the bulk of the distribution deviates significantly from the origin of the coordinate system. Thus, it is usually reasonable to choose $N_{l}$ to be much smaller than $N_{\xi}$, the number of points in the $\xi$ grid.

The mapping between the 2D-finite-difference-grid and finite-difference--Legendre-mode representations can be formulated as a single matrix operation, where a mapping matrix $M\sub{L}$ can be constructed to represent the summation in \Eq{eq:f_leg_modes}. In general, $M\sub{L}$ is not square, but the inverse mapping can be performed by taking the Moore-Penrose pseudo-inverse of $M\sub{L}$ to find the inverse in a least-squares sense. (This only needs to be done once in each \NORSE run.) 
The solution is exact in the sense that norm$\left[M\sub{L}f_{l}(p)-f(p,\xi)\right]$
is of order the round-off error, and the mapping between the two representations can thus be performed to machine precision at very small computational cost. 

The parallel axis is the symmetry axis of the problem. Therefore,
we require that the derivative of the distribution with respect to
$p_{\perp}$ at a fixed $p_{\|}$ must vanish as $p_{\perp}\to0$. This condition must be imposed as a boundary condition at $p_{\|}=0$, but is automatically satisfied for all non-vanishing $p_{\|}$. At $p=p_{\max}$, we impose the Dirichlet condition $f(p\sub{max})=0$ for all $\xi$.

%%%%%%%%%%%%%%%%%%%%%%%%%%%%%%%%%%%%%%%%%%%%%%%%%%%%%%%%%%%%%%%%%%

\subsection{Calculation of potentials}
The Legendre modes of the potentials $\Psi$ can be calculated from the distribution using \Eq{eq:potentials_relations}, which becomes
\begin{equation}
L_{0,l}\Upsilon_{0,l} =f_l,\quad 
L_{2,l}\Upsilon_{1,l}=\Upsilon_{0,l},\quad 
L_{2,l}\Upsilon_{2,l}=\Upsilon_{1,l},\quad 
L_{1,l}\Pi_{0,l}=f_l,\quad 
L_{1,l}\Pi_{1,l}=\Pi_{0,l},
\label{eq:potentials_relations_l}
\end{equation} 
where 
\begin{equation}
L_{a,l}\Psi 
= \gamma^{2}\pdersec{\Psi}{p}
+ \left(
	\frac{2}{p}+3p
  \right)\pder{\Psi}p
+ \left(
	1 - a^2 - \frac{l(l+1)}{p^2}
  \right)\Psi
\end{equation}
is obtained by decomposing the differential operator $L_a$ (described in our coordinate system by Eq.~\ref{eq:L_a_pxi}) into Legendre modes. Inverting Eqs.~\eqref{eq:potentials_relations_l} results in operators which can determine the potentials from an arbitrary $f$, and a block-diagonal sparse matrix for each potential $\Psi$ can be constructed to efficiently calculate $\Psi_l$ from $f_l$ for all $l$ using a single matrix multiplication, in accordance with the discussion in Section~\ref{sec:Performance}. 

The above calculation requires that boundary conditions for the potentials be specified. In general, for a function $\phi(p,\xi)$ to be continuous at $p=0$, its Legendre modes $\phi_l(p)$ must satisfy $\partial\phi_{l}(0)/\partial p=0$ for $l=0$ and $\phi_l(0)=0$ for $l>0$. Boundary conditions at $p=p_{\max}$ can be determined from Eq.~(31) in \Ref{BraamsKarney1989}, which gives explicit expressions for the potentials $\Psi_l$ in terms of weighted integrals over $f_l$. The calculation of these boundary conditions is discussed in \ref{app:potentials_bcs}.

%%%%%%%%%%%%%%%%%%%%%%%%%%%%%%%%%%%%%%%%%%%%%%%%%%%%%%%%%%%%%%%%%%

\subsection{Time advance}

To advance the system in time, we employ a linearly implicit time-advancement scheme based on the first-order backward-Euler method. The scheme avoids the restriction on the time step imposed on explicit methods by the CFL condition, and is straight-forward to implement, as it only requires building and inverting a single matrix in each time step.  Compared to fully implicit methods, however; the time step has to be kept relatively short, and the overall computation time can still be considerable when simulating a long time span. As long as the time step is short enough, good accuracy is achieved, and this simple scheme is sufficient for our purposes. 

The method is formulated as follows. The entire kinetic equation, excluding the time derivative, can in general be written as an operator $\mathbb{O}\{\Psi\{f\},f\}$, where $\Psi$ represents the five potentials $\Upsilon_i$ and $\Pi_i$, which depend on the distribution $f$. In a fully implicit time-advancement scheme, this operator should be evaluated at the next time step $(k+1)$: $\mathbb{O}\{\Psi\{f^{k+1}\},f^{k+1}\}$. If the potentials are instead evaluated based on the distribution at the current time step, $f^k$, $\mathbb{O}$ can be written as a regular matrix operation $\mathbb{O}\{\Psi\{f^k\},f^{k+1}\} = M_{mn}^k f^{k+1}$, where the matrix $M_{mn}^k=M^k(p_m,\xi_n)$ describes a set of linear equations. This makes the time-advancement scheme linearly implicit, and $M^k$  can be explicitly evaluated in each time step and the system solved using standard matrix-inversion techniques. The Backward-Euler method for our problem can then be written as
\begin{equation}
f^{k+1} = f^{k} + \Delta t M^{k} f^{k+1},
\end{equation}
where $\Delta t$ is the time step.

%%%%%%%%%%%%%%%%%%%%%%%%%%%%%%%%%%%%%%%%%%%%%%%%%%%%%%%%%%%%%%%%%%

\subsection{Performance}
\label{sec:Performance}
\NORSE is written in \texttt{Matlab}, using an object-oriented structure. To make efficient use of the \texttt{Matlab} language, care has been taken to formulate the problem in terms of matrix multiplications and avoid loops where they are detrimental to performance. To this end, many parts of the operators of the kinetic equation are pre-calculated to speed up the matrix building in each time step. As an example, the first term of the electron-electron collision operator Eq.~\eqref{eq:Cee_components} at time step $k$ (together with Eq.~\ref{eq:p2_term}) can be written as 
\begin{equation}
\prefactor^{(p^{2})} \pdersec{f^{k+1}}{p} 
= \left(
	\mathbb{\prefactor}_{0}\Upsilon_{0}^k
	+\mathbb{\prefactor}_{2}\Upsilon_{2}^k
	+\mathbb{\prefactor}_{-}\Upsilon_{-}^k
  \right)
  \mathbb{D}_{pp}^{2} f^{k+1},
\label{eq:operator_example}
\end{equation}
where the various operators, defined as
\begin{align}
\mathbb{\prefactor}_{0} & =-\gamma,\qquad
\mathbb{\prefactor}_{2} = 8\gamma, \qquad
\mathbb{\prefactor}_{-} =-2\frac{\gamma^{3}}{p}\mathbb{D}_{p}-\frac{\gamma(1-\xi^{2})}{p^{2}}\mathbb{D}_{\xi\xi}^{2}+2\frac{\gamma\xi}{p^{2}}\mathbb{D}_{\xi} \\
\mathbb{D}_{pp}^{2} & =\pdersec{}p,\qquad\mathbb{D}_{p}=\pder{}p,\qquad\mathbb{D}_{\xi\xi}^{2}=\pdersec{}{\xi},\qquad\mathbb{D}_{\xi}=\pder{}{\xi},
\end{align}
are all independent of $f$, and can thus be pre-calculated. Constructing this part of the linear system in each time step is thus reduced to determining the potentials $\Upsilon_{i}^k$ from $f^k$ and constructing $\prefactor^{(p^{2})} \mathbb{D}^2_{pp}$ in accordance with \Eq{eq:operator_example}, using just a few matrix operations.  

The above algorithm is efficient, making the matrix inversion associated with the solution of the resulting linear system the most costly part of each time step. The overall computational cost can be reduced by approximately a factor of 2 by employing an iterative scheme using the generalized minimal residual method (\texttt{gmres} \cite{SaadSchultz1986}), which is available in \texttt{Matlab} as a standard subroutine. By periodically (every $n\sub{LU}$ time steps) solving the system exactly using LU-factorization, and supplying the L and U factors as preconditioners for the next $n\sub{LU}-1$ steps, \texttt{gmres} converges in just a few iterations if $n\sub{LU}$ is sufficiently small. 

In certain scenarios -- such as where an initial transient requires high temporal resolution, but the subsequent relaxation happens on a significantly longer time scale -- adaptive time-step schemes can be very effective in reducing the computational expense. Such a scenario is for example considered in Section~\ref{sec:Two_Maxwellians}. Here, we use a simple adaptive-time-step scheme based on information about the number of iterations needed for convergence of the \texttt{gmres} algorithm. 
If few \texttt{gmres} iterations are needed for convergence ($n\sub{gmres}<n\sub{opt}$, where $n\sub{opt}$ is some desired optimal number), the change in the distribution in each time step is small, indicating that the step length can be increased. Conversely, the step length should be reduced if $n\sub{gmres}>n\sub{opt}$. 

Employing the techniques discussed above makes the implementation efficient, and moderately sized test cases usually run on a standard laptop in less than a minute. 

%%%%%%%%%%%%%%%%%%%%%%%%%%%%%%%%%%%%%%%%%%%%%%%%%%%%%%%%%%%%%%%%%%%%%%%%%%%%
%%%%%%%%%%%%%%%%%%%%%%%%%%%%%%%%%%%%%%%%%%%%%%%%%%%%%%%%%%%%%%%%%%%%%%%%%%%%
%%%%%%%%%%%%%%%%%%%%%%%%%%%%%%%%%%%%%%%%%%%%%%%%%%%%%%%%%%%%%%%%%%%%%%%%%%%%

\section{Tests and benchmarks}
\label{sec:Tests}
The kinetic equation solved by \NORSE is valid for strongly non-Maxwellian electron distributions, as well as relativistic temperatures and particle energies. In this section, we validate the implementation by comparing to the two limits of arbitrary temperature but weakly distorted distribution (Section \ref{sec:BK_conductivity}), and non-relativistic but fully non-linear distribution (Section \ref{sec:Weng_comparison}). 
However, let us first look at a proof-of-principle scenario, demonstrating both the non-linearity as well as the high-temperature validity of \NORSE. In this section, we will repeatedly make use of the normalized time $\tau = \nu t$ (i.e. the time in units of relativistic-electron collision times) and the normalized distribution $F=f/f_M(p=0)$ (so that if the initial distribution is a Maxwellian, $F$ initially takes the value unity at $p=0$). We will also use the normalized electric-field magnitude $\hat{E}=eE/\me c\nu=E/E\sub{c}$. Throughout the rest of the paper, we will always apply fields with an implicit minus sign, so that electrons will be accelerated towards positive $p_{\|}$. 

%%%%%%%%%%%%%%%%%%%%%%%%%%%%%%%%%%%%%%%%%%%%%%%%%%%%%%%%%%%%%%%%%%%%%%%%%%%%

\subsection{Proof-of-principle non-linear scenario: collisional relaxation of a two-Maxwellian initial state}
\label{sec:Two_Maxwellians}

\begin{figure}
% Figure generated using RelaxingMaxwellians.m
\begin{center}
\includegraphics[width=0.49\textwidth, trim={0cm 0cm 0.0cm 0.0cm},clip]{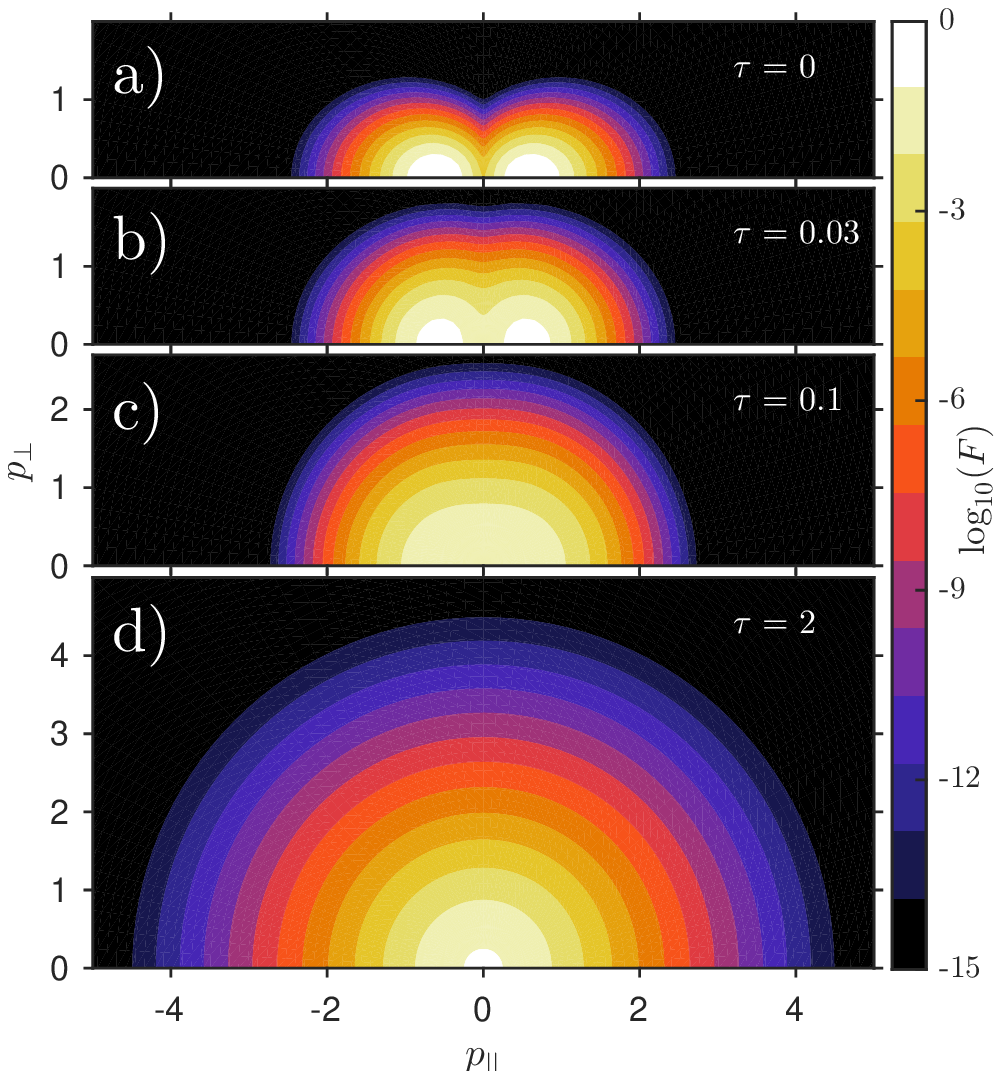}
\includegraphics[width=0.49\textwidth, trim={0cm 0cm 0.0cm 0.0cm},clip]{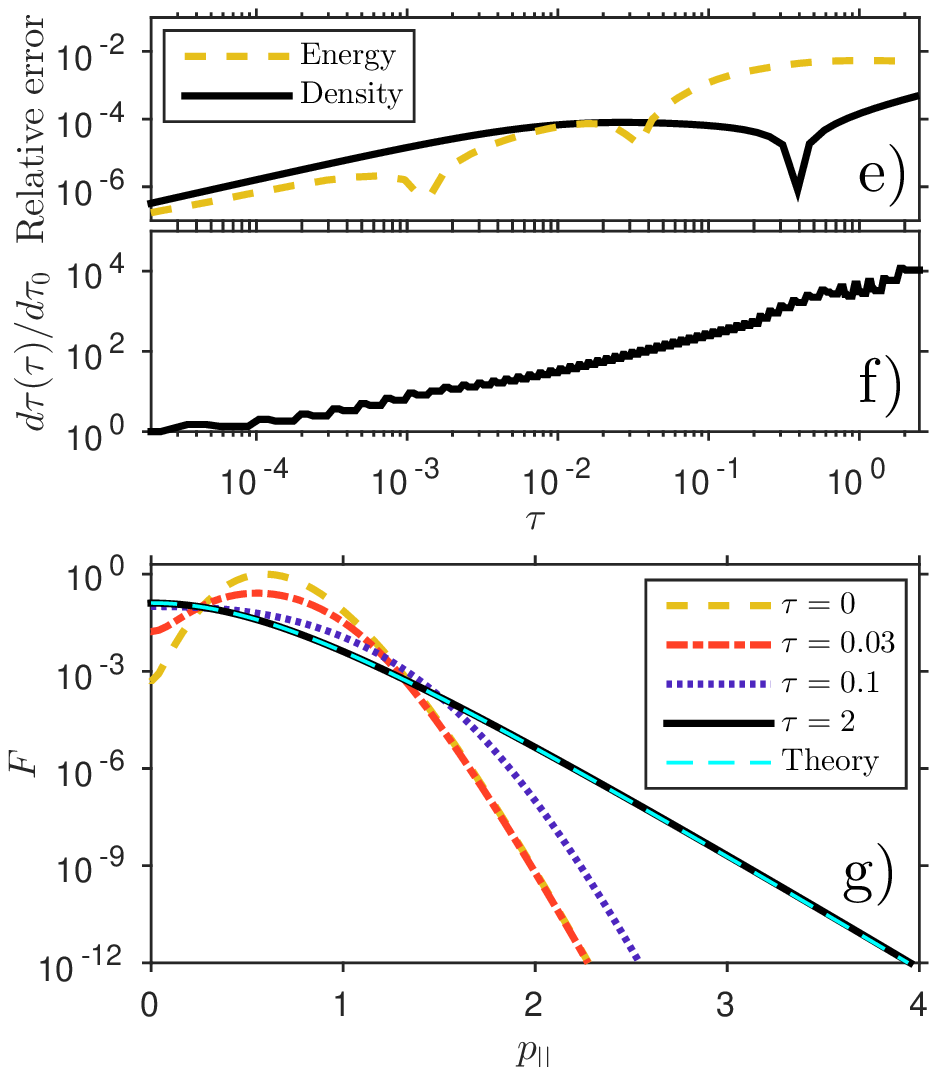}
\caption{Collisional relaxation from a starting distribution consisting of two shifted Maxwellians. Panels a)--d) show the 2D distribution at various times, and panel g) shows corresponding cuts along the positive parallel axis. Panel e) shows the conservation of density and energy and panel f) the time step used, as functions of time. The numerical parameters $N_p=250$, $N_{\xi}=65$, $N_l=25$, $p_{\max}=10$, and the initial time step $d\tau_0=0.001$ were used. A uniform grid was used in $p$, whereas a non-uniform grid giving uniform spacing in the polar angle ($\arccos\xi$) was used for the $\xi$ coordinate.}
\label{fig:TwoMaxwellians}
\end{center}
\end{figure}

In this section we demonstrate the validity of the \NORSE implementation by considering a basic non-linear test case: the collisional relaxation of two initially shifted Maxwellians. A shifted Maxwell-J\"uttner distribution (e.g. the equilibrium distribution with temperature $\Theta\sub{b}$ in a frame boosted by $p\sub{b}$ in the parallel direction, as seen from the stationary frame) takes the form 
\begin{equation}
f\sub{M,b}(\Theta\sub{b},p\sub{b}) 
= \frac{n}{4\pi\Theta\sub{b} K_{2}\bigl(1/\Theta\sub{b}\bigr)}
\exp\left( -\frac{\gamma\sub{b}\gamma-p\sub{b}p_{\|}}{\Theta\sub{b}} \right),
\end{equation} 
with $\gamma\sub{b} = \sqrt{1+p\sub{b}^2}$ and $p_{\|}=p\xi$. We consider two initial Maxwellians, each with a temperature of 10 keV ($\Theta\sub{b} = 0.0196$), and each shifted the equivalent of three thermal speeds ($p\sub{b}=0.59$) along the symmetry axis (in opposite directions). The initial state is depicted in \Fig{fig:TwoMaxwellians}a), with panels b)--d) showing the subsequent evolution of the distribution function. Panel g) shows a cut of the distribution along the positive parallel axis at the same time steps. 
The parameter values $E=0$ and $B=0$ were used, and to isolate the behavior of the non-linear electron self-collision operator, a pure electron plasma was assumed ($\zeff =0$). The number density of each Maxwellian in its rest frame was set to $n=10^{19}\,$m$^{-3}$, resulting in a total initial number density of $n\sub{tot}=2\gamma\sub{b}n=2.326\power{19}\,$m$^{-3}$. The expected final-state Maxwellian (cyan, thin dashed line in panel g) has a temperature of 61.3 keV, which can be calculated by equating \Eq{eq:W_Maxwellian} (with $n\to n\sub{tot}$) with the combined energy content in the two shifted Maxwellians:
\begin{equation}
	W\sub{tot} = 2 W\sub{b},\qquad
	W\sub{b} = \gamma\sub{b}^2 W 
			 + \me c^2 n
			    \left[
			    		\gamma\sub{b}(\gamma\sub{b}-1)
			    		+ (\gamma\sub{b}^2-1)\Theta\sub{b}
			    \right]
\end{equation}
(with $W$ given by \ref{eq:W_Maxwellian}), and solving for $\Theta$. The final equilibrium state shows excellent agreement with the theoretical prediction.

The relative error (compared to the initial value) in the density and energy contents of the \NORSE solution are shown in \Fig{fig:TwoMaxwellians}e) as functions of time. For the numerical parameters used, the density is conserved to within 0.05\%, whereas the relative error in energy saturates at the 0.5\% level. Figure~\ref{fig:TwoMaxwellians}f) shows the time step used by the adaptive-time-step scheme, normalized to the initial time step. In this particular case, the scheme is very effective since the time evolution involves an initial transient followed by a comparatively slow asymptotic relaxation. The final time step was approximately $10^4$ times longer than the initial time step, and a total of 312 time steps were used (as opposed to $\sim 4\power{5}$ had the initial time step been used throughout the entire calculation).

%%%%%%%%%%%%%%%%%%%%%%%%%%%%%%%%%%%%%%%
%%%%%%%%%%%%%%%%%%%%%%%%%%%%%%%%%%%%%%%

%%%%%%%%%%%%%%%%%%%%%%%%%%%%%%%%%%%%%%%%%%%%%%%%%%%%%%%%%%%%%%%%%%%%%%%%%%%%

\subsection{Weak-electric-field limit: conductivity for relativistic temperatures}
\label{sec:BK_conductivity}
In \Ref{BraamsKarney1989}, Braams \& Karney use the relativistic electron-electron collision operator to calculate the plasma conductivity for a wide range of temperatures. The operator is linearized around a stationary Maxwellian, and the zeroth and first Legendre modes are calculated numerically as an initial-value problem. The results are compiled in their Table 1, which contains normalized conductivities for $\Theta\in [0,100]$ (recall that $\Theta=1$ corresponds to $T=\me c^2\simeq 511\,$keV) and $\zeff\in [0,\infty]$. The unit used is 
\begin{equation}
	\bar{\sigma} = 
	\frac{\sqrt{e\me}\lnL \zeff}{4\pi\epso^2 T^{3/2}} 
	\frac{j}{E},
\end{equation}
where $j$ is the current density.

%%%%%%%%%%%%%%%%%%%%%%%%%%%%%%%%%%%%%%%
\begin{figure}
% Figure generated using Test.m
\begin{center}
\includegraphics[width=0.49\textwidth, trim={0cm 0cm 0.0cm 0.0cm},clip]{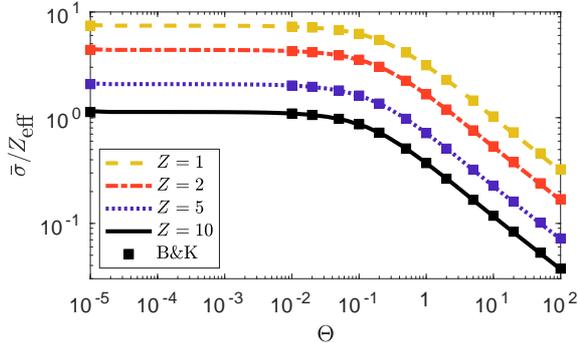}
\caption{Normalized conductivity in \NORSE (lines) for various temperatures and plasma compositions. Data points from Table 1 of \Ref{BraamsKarney1989} are also shown (squares). The electric field corresponded to $E=10^{-3}E\sub{D}$\mod{, and} $n=5\power{19}\,$m$^{-3}$ and $B=0$ \mod{were used}.}
\label{fig:BKFig1}
\end{center}
\end{figure}
%%%%%%%%%%%%%%%%%%%%%%%%%%%%%%%%%%%%%%%

To demonstrate that our implementation reproduces the above results, we similarly calculate the conductivity of a quasi-steady-state distribution found by evolving the system \mod{from a Maxwellian initial state using a constant electric field corresponding to $E=10^{-3}E\sub{D}$. H}owever, we make no simplification to the collision operator and retain adequate resolution in $\xi$ to accurately resolve the distribution function in 2D momentum space. \Fig{fig:BKFig1} overlays \NORSE results with the data in \Ref{BraamsKarney1989} for $\zeff=1,2,5$ and $10$, and all tabulated temperatures. Excellent agreement is seen for all parameters. \mod{In the figure, the data points at $\Theta=0$ in \Ref{BraamsKarney1989} are compared to \NORSE runs with $\Theta=10^{-5}$, however all temperatures $\Theta<10^{-3}$ give good agreement as the obtained values of $\bar{\sigma}$ are essentially independent of the temperature in this range.}

%%%%%%%%%%%%%%%%%%%%%%%%%%%%%%%%%%%%%%%%%%%%%%%%%%%%%%%%%%%%%%%%%%%%%%%%%%%%

\subsection{Non-relativistic limit: highly anisotropic distributions}
\label{sec:Weng_comparison}
Several codes exist that solve the non-relativistic kinetic equation using non-linear collision operators. To validate the non-linear aspect of \NORSE, we will compare to conductivities reported by Weng et al. in \Ref{Weng2008}. In their Fig.~3, conductivities as functions of time are presented for electric fields as strong as the Dreicer field $E\sub{D}$, leading to highly distorted distributions. Results are shown for $E/E\sub{D}=0.01,0.1$ and 1, with $\zeff=1$ and $B=0$.

\begin{figure}
% Figure generated using Test.m
\begin{center}
\includegraphics[width=0.49\textwidth, trim={0cm 0cm 0.80cm 0.45cm},clip]{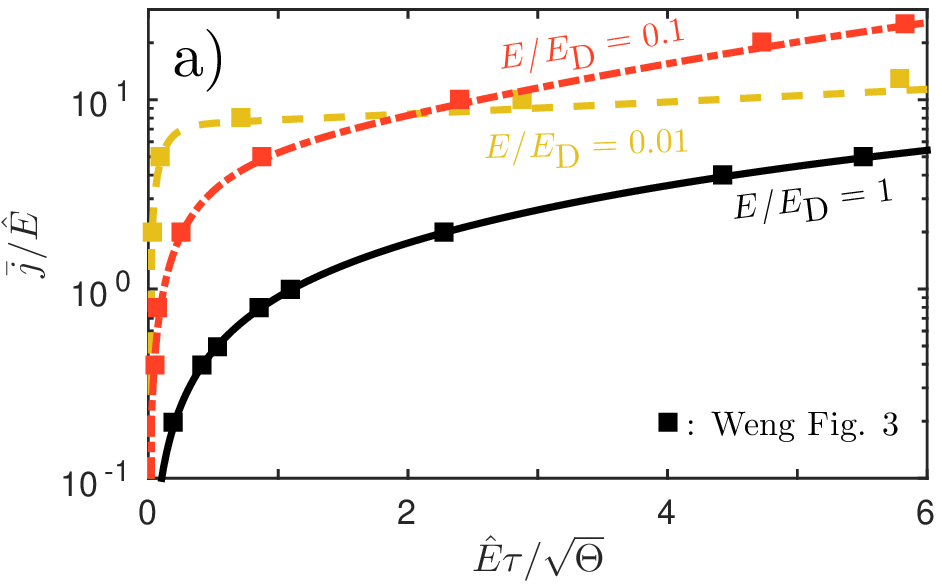}
\includegraphics[width=0.49\textwidth, trim={0cm 0cm 0.80cm 0.45cm},clip]{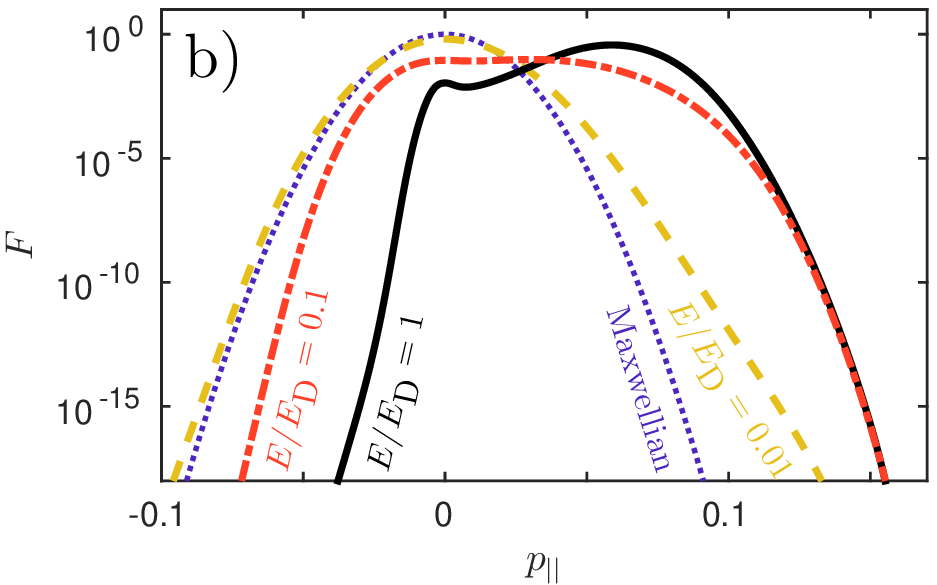}
\caption{a) Normalized conductivity in \NORSE (lines) as a function of time for various E-field strengths. Data points extracted from Fig.~3 of \Ref{Weng2008} are also shown (squares). b) Cuts in the parallel direction through the final distributions in a). The numerical parameters were $N_p=300$, $N_\xi=55$, $N_l=15$, $p\sub{max}=0.3$, using 2000, 400 and 300 time steps for $E/E\sub{D}=0.01,\,0.1$ and 1, respectively.} 
\label{fig:WengFig3}
\end{center}
\end{figure}

Figure \ref{fig:WengFig3}a) reproduces the results in Fig.~3 of \Ref{Weng2008} using \NORSE. The units used are those of the original figure: conductivities are given as $\bar{j}/\hat{E}$, with $\bar{j}=j\zeff/nec\Theta^{3/2}$ a normalized current density, and the time unit used is $\hat{E}\tau/\sqrt{\Theta}$. The parameters $\Theta=1\power{-4}$ (corresponding to a temperature of $T=51\,$eV) and $n=5\power{19}\,$m$^{-3}$ were used in \NORSE. Data points extracted from the figure in \Ref{Weng2008} are included for comparison. The agreement is very good in general, demonstrating that \NORSE behaves as expected also for highly non-linear distributions. The error is somewhat larger (and systematic) for the weakest electric field, however in this case, numerical heating in the results of \Ref{Weng2008} cannot be ruled out \cite{Weng2016}. The final distributions for the three field strengths are shown in \ref{fig:WengFig3}b). As can be seen, the distributions deviate strongly from the initial Maxwellian for the two higher field strengths, and even at the weakest field, a substantial tail of runaway electrons is produced. For the strongest field, a small ``bump'' in the distribution is seen at $p_{\|}=0$, which indicates that electron-ion collisions are strong enough to ``capture'' a sub-population of the electrons, despite the strong accelerating electric field.

%%%%%%%%%%%%%%%%%%%%%%%%%%%%%%%%%%%%%%%
%%%%%%%%%%%%%%%%%%%%%%%%%%%%%%%%%%%%%%%

%%%%%%%%%%%%%%%%%%%%%%%%%%%%%%%%%%%%%%%%%%%%%%%%%%%%%%%%%%%%%%%%%%%%%%%%%%%%
%%%%%%%%%%%%%%%%%%%%%%%%%%%%%%%%%%%%%%%%%%%%%%%%%%%%%%%%%%%%%%%%%%%%%%%%%%%%
%%%%%%%%%%%%%%%%%%%%%%%%%%%%%%%%%%%%%%%%%%%%%%%%%%%%%%%%%%%%%%%%%%%%%%%%%%%%

\section{Application: Runaway electrons in fusion plasmas}
\label{sec:Application--runaway}
As an example of how \NORSE can be used to provide new physical insight, we consider the case of runaway-electron generation in magnetically confined fusion plasmas. Under the influence of strong electric fields, electrons quickly accelerate to relativistic speeds since the friction force they experience decreases with increasing velocity. The localized heat loads associated with the eventual loss of fast-electron confinement constitute a serious threat to the plasma-facing components of fusion reactors. The energy and current carried by the runaway electrons, and thus the potential for damage, increase the larger the distortion of the electron distribution. Unlike existing tools such as \texttt{LUKE} \cite{Decker2004} and \CODE\cite{CODEPaper2014,Stahl2016}, \NORSE can be used to study the cases of highest runaway-electron growth rate.

%%%%%%%%%%%%%%%%%%%%%%%%%%%%%%%%%%%%%%%%%%%%%%%%%%%%%%%%%%%%%%%%%%%%%%%%%%%%

\subsection{Runaway region of momentum space}
\mod{What constitutes a runaway particle can be defined in several ways. The definitions usually employed in theoretical works, as well as many numerical tools, assume the distribution to be close to Maxwellian and are therefore not directly applicable in our context \cite{Dreicer1959,Smith2005}. In addition, it has been pointed out that synchrotron radiation reaction may have a significant impact on the runaway region \cite{MartinSolis1998,Stahl2015,Aleynikov2015,Liu2016}, however the commonly used definitions only account for collisional friction. We will define a runaway region based on particle trajectories in momentum space \cite{Smith2005}, neglecting the effect of diffusion but allowing for arbitrary electron distributions as well as synchrotron radiation reaction.}

\mod{For an arbitrary electron distribution, the lower boundary of the runaway region (the \emph{separatrix})} can be obtained by considering the forces that affect a test particle:
\begin{align}
	\frac{\d p}{\d t} &= F\sub{E}^p - F\sub{C}^p - F\sub{S}^p 
		= \frac{eE}{\me c}\xi 
		+ \alpha \gamma \pder{\Pi}{p}
		+ \frac{\gamma p (1-\xi^2)}{\tau\sub{r}}, 
		\label{eq:sum_of_forces}\\
	\frac{\d \xi}{\d t} &= F\sub{E}^{\xi} - F\sub{C}^{\xi} - F\sub{S}^{\xi}
		= \frac{eE}{\me c} \frac{1-\xi^2}{p}
		+ \alpha \frac{1-\xi^2}{\gamma p^2} \pder{\Pi}{\xi}
		- \frac{\xi (1-\xi^2)}{\gamma \tau\sub{r}},
\end{align}
where the expressions for the force associated with the electric field, $F\sub{E}^i$, the collisional electron-electron friction $F\sub{C}^i$ and the synchrotron radiation-reaction force $F\sub{S}^i$ are taken from Eqs.~\eqref{eq:kinetic_eq}, \eqref{eq:F} and \eqref{eq:synch_force_components}, respectively. \mod{Asymptotically, particles on the separatrix neither end up in the bulk population nor reach arbitrarily high energies, but instead settle at the point $p\sub{c}$ of parallel force balance at $\xi=1$ (in the absence of diffusion).} $p\sub{c}$ can be determined from $\d p/\d t=0$ at $\xi=1$, since the separatrix becomes purely perpendicular to the parallel axis as $\xi\to 1$. The separatrix is then traced out by numerically integrating the above equations from $\xi=1$ to $\xi=-1$. In the \mod{limit of non-relativistic temperature ($\Theta \ll 1$), small departure from a Maxwellian, and $B=0$, the result agrees with the standard expression \cite{Smith2005}}. To ensure consistency with the distribution, the separatrix in \NORSE is calculated in each time step.

%%%%%%%%%%%%%%%%%%%%%%%%%%%%%%%%%%%%%%%%%%%%%%%%%%%%%%%%%%%%%%%%%%%%%%%%%%%%

\subsection{Distortion-induced transition to electron slide-away}

\begin{figure}
\begin{center}
\includegraphics[height=0.4\textheight, trim={0cm 0cm 0.0cm 0.0cm},clip]{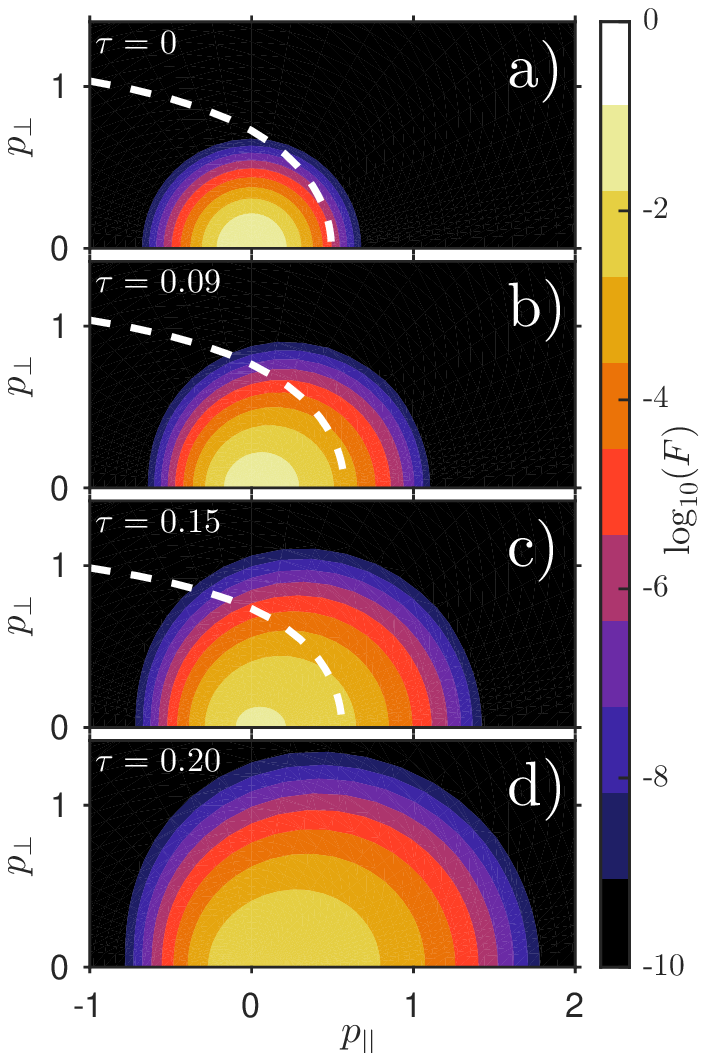}
\includegraphics[height=0.4\textheight, trim={0cm 0cm 0.0cm 0.0cm},clip]{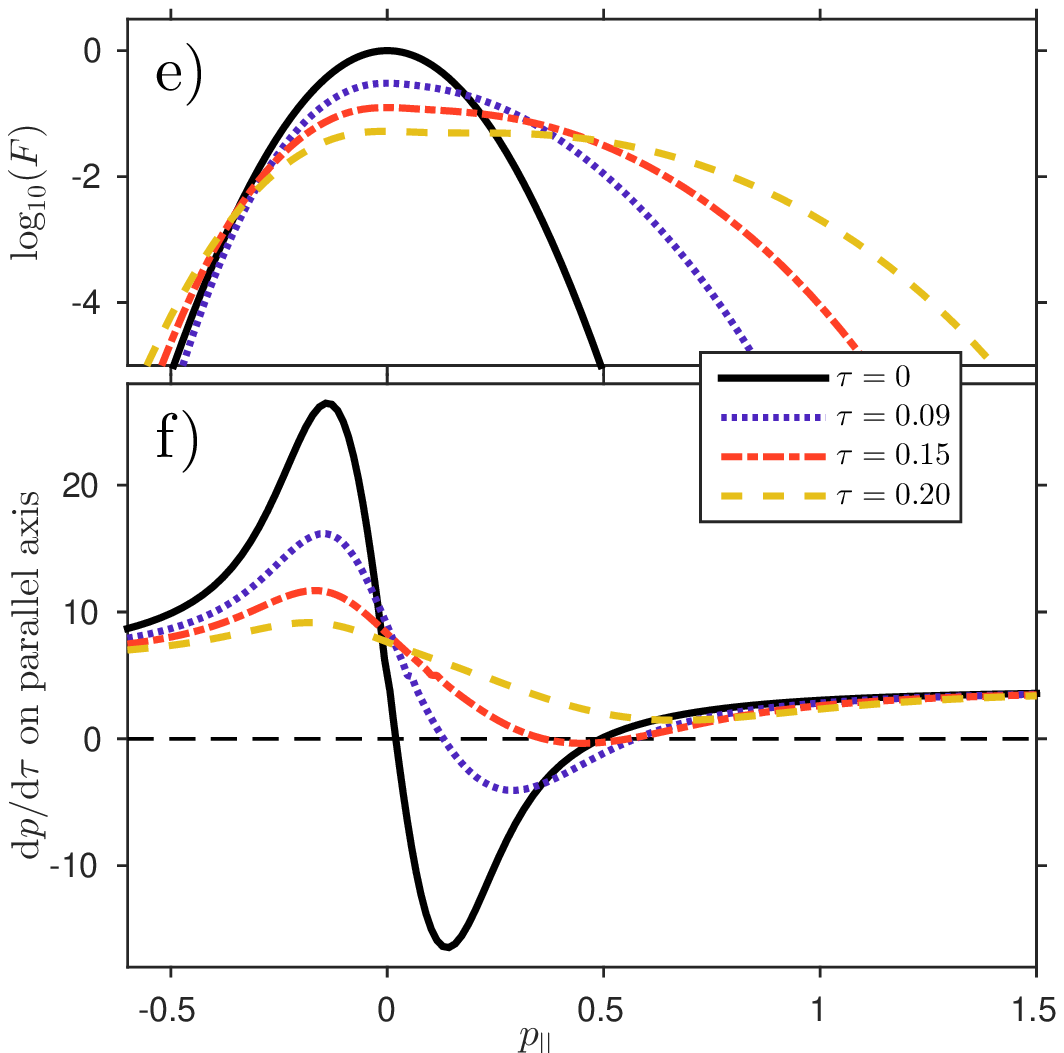}
\caption{a)--d) Contour plots and e) cuts along the parallel axis of the distribution at different times during a \NORSE run. In a)--c), the white dashed lines are the separatrices defining the lower boundary of the runaway region. f) Sum of forces (neglecting diffusion) on the parallel axis. The physical parameters were $\Theta=0.01$ ($T=5.11\,$keV), $n=5\power{19}$m$^{-3}$, $\zeff=1$, $\hat{E}=5$ (corresponding to $E/E\sub{D,0}=0.05$ and $E=0.22\,$V/m), $B=0$, and  $N_p=100$, $N_{\xi}=35$, $N_l=5$, $p_{\max}=3.54$, and $d\tau=2.8\power{-4}$ were used.}
\label{fig:sum_of_forces}
\end{center}
\end{figure}

For electric fields stronger than approximately $E\sub{sa}=0.215 E\sub{D}$, all electrons in a Maxwellian distribution experience net acceleration, since the field overcomes the maximum of the collisional friction force. This is known as \emph{electron slide-away} \mod{\cite{Dreicer1959,Coppi1976}}. However, in a non-linear treatment, the condition for slide-away can in principle be modified since the collisional friction depends on the shape of the electron distribution. The distortion of the distribution associated with a moderately strong electric field turns out to have a large effect on the effective Dreicer field at which the transition to the slide-away regime occurs. This is illustrated in Fig.~\ref{fig:sum_of_forces}, which shows the distribution at several time steps, as well as the separatrix and the force balance (neglecting diffusion) as a function of $p$ at $\xi=1$. In the figure, the distribution is evolved under a constant electric field which initially corresponds to 5\% of the Dreicer field (\mod{$E=0.05 E\sub{D,0}\approx0.23E\sub{sa,0}$, with $E\sub{D,0}$ and $E\sub{sa,0}$ the Dreicer and slide-away fields at the initial temperature}). The distribution quickly becomes distorted, and soon after $t=0.15\tau$ the slide-away regime is reached. This can be seen in Fig.~\ref{fig:sum_of_forces}f), where the sum of forces (Eq.~\ref{eq:sum_of_forces}) becomes positive everywhere on the parallel axis, indicating that the electric field at that time corresponds to the instantaneous slide-away field, $E=E\sub{sa}(t)$. No separatrix therefore exists for later times (see Fig.~\ref{fig:sum_of_forces}d).

%%%%%%%%%%%%%%%%%%%%%%%%%%%%%%%%%%%%%%%
% Figure generated using DoNORSERun_separatrix_fig_for_paper.m
%%%%%%%%%%%%%%%%%%%%%%%%%%%%%%%%%%%%%%%

An important effect of the electric field is to quickly heat the bulk of the distribution, and this turns out to be the main cause of the induced transition to slide-away. Since (neglecting the weak dependence on $\lnL$) the Dreicer field $E\sub{D}\sim 1/T$, an increase in temperature lowers the effective Dreicer field and thus the threshold for electron slide-away. An approximate effective temperature $T\sub{eff}$ can be estimated from the energy moment $W$ of the \NORSE distribution by solving \Eq{eq:W_Maxwellian} for $\Theta$. The effective Dreicer field can then easily be evaluated. 

\begin{figure}
\begin{center}
\includegraphics[width=0.70\textwidth, trim={0cm 0cm 0.0cm 0.0cm},clip]{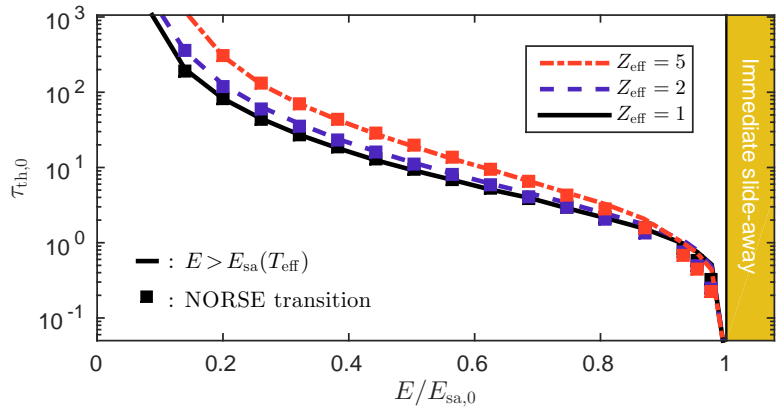}
\caption{Time to transition to slide-away as a function of electric-field strength for various values of $\zeff$. Times calculated from the effective temperature of \NORSE distributions are indicated by lines, while the actual times obtained in \NORSE (based on force balance) are indicated by squares. Here, a subscript 0 indicates an initial value, and $\tau_{\mathrm{th},0}=\tau/(2\Theta_0)^{3/2}$ denotes the initial thermal-electron collision time. The physical parameters were $\Theta_0=1\power{-4}$, $n=10^{19}$m$^{-3}$, $B=0$.}
\label{fig:slide-away_transition}
\end{center}
\end{figure}

Figure \ref{fig:slide-away_transition} highlights the importance of the heating effect by showing the time to a transition to slide-away under constant electric fields of various strengths (starting from an equilibrium distribution). Lines denote when the effective temperature becomes such that $E>E\sub{sa}(T\sub{eff})$, whereas squares denote the actual transition in \NORSE, calculated from the force balance. The agreement between these two values is very good in the entire range of electric-field values, demonstrating that the bulk heating is the dominant effect in the modification of the slide-away threshold. Only at fields very close to $E\sub{sa,0}$ does the values obtained using the effective temperature noticeably overestimate the time to transition, indicating that here, other effects start to become important as well. The figure also shows that the process leading to a transition to slide-away is quick, also for relatively weak fields. At $E/E\sub{sa,0}=0.3$ ($E/E\sub{D,0}\approx 0.065$), the transition happens around 30 thermal collision times for $\zeff=1$, and at $E/E\sub{sa,0}=0.1$ ($E/E\sub{D,0}\approx 0.022$), the corresponding figure is 500.

In practice, various processes can lead to heat losses, as previously noted. This can partially or entirely offset the heating caused by the electric field, and in many situations the modification to the slide-away threshold may not be as dramatic as demonstrated here. In addition, a feedback mechanism commonly exists between the accelerating electric field and the distribution (through changes in the plasma current). In such scenarios, a reduction in the electric field may be induced due to the changes in the distribution before these have become too extensive, thus limiting the distortion and potentially avoiding a transition to slide-away altogether. 

%%%%%%%%%%%%%%%%%%%%%%%%%%%%%%%%%%%%%%%
% Figure generated using Test.m
%%%%%%%%%%%%%%%%%%%%%%%%%%%%%%%%%%%%%%%

%%%%%%%%%%%%%%%%%%%%%%%%%%%%%%%%%%%%%%%%%%%%%%%%%%%%%%%%%%%%%%%%%%%%%%%%%%%%
%%%%%%%%%%%%%%%%%%%%%%%%%%%%%%%%%%%%%%%%%%%%%%%%%%%%%%%%%%%%%%%%%%%%%%%%%%%%
%%%%%%%%%%%%%%%%%%%%%%%%%%%%%%%%%%%%%%%%%%%%%%%%%%%%%%%%%%%%%%%%%%%%%%%%%%%%

\section{Conclusions}

The study of energetic-electron populations in plasmas has long been of interest, but when considering relativistic particles in kinetic simulations, the work has so far been restricted to linearized treatments of the Fokker-Planck collision operator. In this paper, we remove that limitation by introducing a new efficient computational tool (\NORSE) which includes the fully non-linear relativistic collision operator in the differential form developed by Braams \& Karney, as well as electric-field acceleration and synchrotron-radiation reaction. A 2D non-uniform finite-difference grid is used to represent momentum space, however when evaluating the five relativistic potentials (analogous to the two Rosenbluth potentials in the non-relativistic case), a mixed finite-difference--Legendre-mode representation is used since the potentials are given by simple 1D integrals in a Legendre-mode decomposition. The system is evolved using a linearly implicit time-advancement scheme, and a simple method for adapting the time step during runtime has been implemented. \NORSE has been successfully benchmarked in both the relativistic-weak-field and non-relativistic--non-linear limits. 

As an application, we have used \NORSE to investigate scenarios relevant to the study of runaway electrons in magnetic-confinement-fusion plasmas. We find that the quick heating of the bulk associated with the application of medium to high-strength electric fields (compared to the Dreicer field) leads to a transition to the electron slide-away regime, despite the $E$ field being weaker than the threshold value $E\sub{sa}=0.215E\sub{D}$ for the initial distribution. The time scale for this transition is relatively short, ranging from a few to a few hundred thermal collision times for $E$ fields in the range $E/E\sub{sa} > 0.1$. These effects cannot be consistently captured in a linearized treatment, and this example thus illustrates that \NORSE opens new avenues of investigation into the dynamics of relativistic electrons in plasmas.

%%%%%%%%%%%%%%%%%%%%%%%%%%%%%%%%%%%%%%%%%%%%%%%%%%%%%%%%%%%%%%%%%%%%%%%%%%%%
%%%%%%%%%%%%%%%%%%%%%%%%%%%%%%%%%%%%%%%%%%%%%%%%%%%%%%%%%%%%%%%%%%%%%%%%%%%%
%%%%%%%%%%%%%%%%%%%%%%%%%%%%%%%%%%%%%%%%%%%%%%%%%%%%%%%%%%%%%%%%%%%%%%%%%%%%

\section*{Acknowledgements}
The authors would like to thank E.~Hirvijoki for his initial work on the Braams \& Karney collision operator, S.-M.~Weng for his helpful assistance, and I.~Pusztai, S.~Newton, T.~DuBois and G.~Wilkie for constructive discussions. This work was supported by the Swedish Research Council (Dnr. 2014-5510), the European Research Council (ERC-2014-CoG grant 647121), and the Knut and Alice Wallenberg Foundation (Dnr. KAW 2013.0078). AS would also like to acknowledge travel support from Adlerbertska Forskningsstiftelsen.

%%%%%%%%%%%%%%%%%%%%%%%%%%%%%%%%%%%%%%%%%%%%%%%%%%%%%%%%%%%%%%%%%%%%%%%%%%%%
%%%%%%%%%%%%%%%%%%%%%%%%%%%%%%%%%%%%%%%%%%%%%%%%%%%%%%%%%%%%%%%%%%%%%%%%%%%%
%%%%%%%%%%%%%%%%%%%%%%%%%%%%%%%%%%%%%%%%%%%%%%%%%%%%%%%%%%%%%%%%%%%%%%%%%%%%

\appendix

\section{Derivation of electron-electron collision operator in $(p,\xi)$ coordinates}
\label{app:Cee_der}

In our coordinate system, the non-zero components of the metric are
\begin{equation}
g_{pp} =1,\qquad\qquad 
g_{\xi\xi}=\frac{p^{2}}{1-\xi^{2}},\qquad\qquad g_{\varphi\varphi}=p^{2}(1-\xi^{2}),
\end{equation}
with $g^{ii}=1/g_{ii}$. Note also that the position vector is just
$\mathbf{p}=p\,\ep$ (with $\ep$ the unit vector along $\mathbf{p}$), since the coordinate system is spherical. Using this and some algebra, we can write the terms in the parenthesis of \Eq{eq:Cee_short} as
\begin{align}
\mathbb{D}\cdot\pder f{\mathbf{p}}  
& =\left[
 	 \left(
 	 	\gamma^{3}\pdersec{\Umin}p
 	 	+\gamma p\pder{\Umin}p
 	 	-\gamma\Uplus
 	 \right)\pder{f}{p}
 	 +\frac{\gamma(1-\xi^{2})}{p^{2}}
 	   g_{\xi\xi}D_{p\xi}(\Umin)\pder f{\xi}
   \right]\ep\nonumber \\
& +\left[
	  \gamma D_{p\xi}(\Umin)\pder{f}{p}
	  +\frac{1-\xi^{2}}{\gamma p^{2}}
	  \left(
		g_{\xi\xi}D_{\xi\xi}(\Umin)
		+p\pder{\Umin}p-\Uplus
	  \right)\pder f{\xi}
	\right]\exi
\label{eq:D_dfdp}
\end{align}
and 
\begin{equation}
\mathbf{F} f
=\gamma\pder{\Pi}p f\mathbf{e}_{p}
+\frac{1-\xi^{2}}{\gamma p^{2}}\pder{\Pi}{\xi} f \mathbf{e}_{\xi},\label{eq:F}
\end{equation}
where 
\begin{align}
D_{\xi\xi}(\Umin) 
&=\frac{1-\xi^{2}}{p^{4}}
	\left[
		(1-\xi^{2})\pdersec{\Umin}{\xi}
		-\xi\pder{\Umin}{\xi}
		+p\pder{\Umin}p
	\right],\nonumber \\
D_{\varphi\varphi}(\Umin) 
&=\frac{1}{p^{4}(1-\xi^{2})}
	\left[
		p\pder{\Umin}p
		-\xi\pder{\Umin}{\xi}
	\right],\label{eq:2nd_der_Ds}\\
D_{p\xi}(\Umin) = D_{\xi p}(\Umin) 
&=\frac{1-\xi^{2}}{p^{4}}
	\left[
		p^{2}\frac{\partial^{2}\Umin}{\partial p\partial\xi}
		-p\pder{\Umin}{\xi}
	\right]\nonumber 
\end{align}
come from the expression for $\partial^2\Umin/\partial\mathbf{p}\partial\mathbf{p}$ in the operator $\mathbb{L}$.

Writing out \Eq{eq:Cee_short} in components, we get
\begin{equation}
C\sub{ee}\{f\}
=\alpha
	\left[
		\frac{1}{p^{2}}\pder{}p
			\left(p^{2}
				\left[
					\mathbb{D}\cdot\pder{f}{\mathbf{p}}
					-\mathbf{F}f
				\right]^{p}
			\right)
	   +\pder{}{\xi}
		\left(
			\left[
				\mathbb{D}\cdot\pder{f}{\mathbf{p}}
				-\mathbf{F}f
			\right]^{\xi}
		\right)
	\right]
\equiv \alpha 
	\left(
		A^p + A^{\xi}
	\right),
\end{equation}
where the superscripts $p$ and $\xi$ denote the corresponding vector components. Carrying out the differentiations, we find the $p$-term to be
\begin{align}
p^2 A^p 
&=
	\left(
		\gamma p^{2}(8\Utwo-\Uzero)		
		-2\gamma^{3}p\pder{\Umin}{p}
		-\gamma(1-\xi^{2})\pdersec{\Umin}{\xi}
		+2\gamma\xi\pder{\Umin}{\xi}
	\right)\pdersec{f}{p} \nonumber \\
&+\left(\ 
	\left[
		\frac{p^{3}}{\gamma}+2\gamma p
	\right](8\Utwo-\Uzero)
	+\gamma p^{2}
	\left[
		-16\pder{\Utwo}p
		+6\pder{\Uone}p
		-\pder{\Uzero}p
	\right]
	-2\gamma^{3}\pder{\Umin}{p}	
  \right. \nonumber \\
& \quad\quad
  \left.
  	-2\gamma^{3}p\pdersec{\Umin}{p}
  	-\frac{p(1-\xi^{2})}{\gamma}\pdersec{\Umin}{\xi}
  	-\gamma(1-\xi^{2})\frac{\partial^{3}\Umin}{\partial p\partial\xi^{2}}
  	+2\frac{p\xi}{\gamma}\pder{\Umin}{\xi}
  	+2\gamma\xi\frac{\partial^{2}\Umin}{\partial p\partial\xi}\ 
  \right)\pder{f}{p} \\
&+ \gamma p^{2}D_{p\xi}(\Umin)\pder{^{2}f}{p\partial\xi}
	+\left[
		\left(
			\frac{p^{3}}{\gamma}+2\gamma p
		\right) D_{p\xi}(\Umin)
		+\gamma p^{2}\pder{D_{p\xi}(\Umin)}{p}
	\right]\pder{f}{\xi} \nonumber \\
&- \gamma p^{2} \pder{\Pi}{p} \pder{f}{p}
	-\left[
		\left(
			\frac{p^{3}}{\gamma}+2\gamma p
		\right)\pder{\Pi}p
		+\gamma p^{2} \pdersec{\Pi}{p}
	\right]f, \nonumber
\end{align} 
whereas the $\xi$-term becomes
\begin{align}
A^{\xi} 
&= \gamma D_{p\xi}(\Umin)\pder{^{2}f}{p\partial\xi}
   +\gamma\pder{D_{p\xi}(\Umin)}{\xi}\pder fp
   +\frac{1-\xi^{2}}{\gamma p^{2}}
   		\left(
   			g_{\xi\xi}D_{\xi\xi}(\Umin)
   			+p\pder{\Umin}{p}
   			-\Uplus
   		\right)\pdersec{f}{\xi}\nonumber \\
&+ \left[
		-2\frac{\xi}{\gamma p^{2}}
			\left(
				g_{\xi\xi}D_{\xi\xi}(\Umin)
				+p\pder{\Umin}p
				-\Uplus
			\right)
		+\frac{1-\xi^{2}}{\gamma p^{2}}
			\left(
				\pder{\left(g_{\xi\xi}D_{\xi\xi}(\Umin)\right)}{\xi}
				+p\pder{^{2}\Umin}{p\partial\xi}
				-\pder{\Uplus}{\xi}
			\right)
	\right]\pder{f}{\xi} \\
&- \frac{1-\xi^{2}}{\gamma p^{2}}\pder{\Pi}{\xi}\pder{f}{\xi}
	+\left[
		2\frac{\xi}{\gamma p^{2}}\pder{\Pi}{\xi}
		-\frac{1-\xi^{2}}{\gamma p^{2}}\pdersec{\Pi}{\xi}
	\right]f. \nonumber
\end{align}
Combining and re-grouping the terms according to the derivative of $f$, we arrive at the final expressions \eqref{eq:p2_term}-\eqref{eq:f_term}. In obtaining some of these results, we have used the differential operator $L_a$, which can be written as
\begin{equation}
L_{a}\Psi 
= \gamma^{2}\pdersec{\Psi}p
+ \left(
	\frac{2}{p}+3p
  \right)\pder{\Psi}p
+ \frac{(1-\xi^{2})}{p^{2}}\pdersec{\Psi}{\xi}
- \frac{2\xi}{p^{2}}\pder{\Psi}{\xi}
+ \left( 1-a^{2} \right)\Psi,
\label{eq:L_a_pxi}
\end{equation}
together with $L_2\Upsilon_\pm = 4\Uone \pm \Uzero$, to remove third order derivatives.

%%%%%%%%%%%%%%%%%%%%%%%%%%%%%%%%%%%%%%%%%%%%%%%%%%%%%%%%%%%%%%%%%%%%%%%%%%%%
%%%%%%%%%%%%%%%%%%%%%%%%%%%%%%%%%%%%%%%%%%%%%%%%%%%%%%%%%%%%%%%%%%%%%%%%%%%%
%%%%%%%%%%%%%%%%%%%%%%%%%%%%%%%%%%%%%%%%%%%%%%%%%%%%%%%%%%%%%%%%%%%%%%%%%%%%

\section{Boundary conditions for the potentials $\Upsilon_i$ and $\Pi_i$ at $p=p_{\max}$}
\label{app:potentials_bcs}
To calculate the potentials $\Upsilon_{i,l}$ and $\Pi_{i,l}$ (or collectively $\Psi_l$) from $f_l$, boundary conditions at $p=p_{\max}$ must be specified. These can be determined from Eq.~(31) in \Ref{BraamsKarney1989}, which for the final grid point becomes
\begin{equation}
\Psi_{l}(p_{\max}) 
	= \int_0^{p_{\max}} \!\! N_{l,*}(p_{\max},p')\frac{p'^2}{\gamma'}f_l(p')\dep',
\end{equation}
where the place holder $*$ denotes a set of indices, distinct for each potential. These indices -- which in the notation of \Ref{BraamsKarney1989} specify the order of differential operators $L_a$ to apply to obtain a given potential from $f$ (cf. Eq.~\ref{eq:potentials_relations}) -- are 
\begin{equation}
\Uzero: 0,\qquad 
\Uone: 02,\qquad
\Utwo: 022, \qquad
\Pzero: 1, \qquad
\Pone: 11.
\end{equation}
The quantity $N_{l,*}$ is defined as 
\begin{equation}
N_{l,*}(p,p') =
	\begin{cases}
		\hfill y_{l,a}(p) j_{l,a}(p'), 
			& \text{if } *=a,	\\
		\hfill y_{l,a}(p) j_{l,aa'}(p')
		+ y_{l,aa'}(p) j_{l,a'}(p'),  
			& \text{if } *=aa', \\
		y_{l,a}(p) j_{l,aa'a''}(p') 
		+ y_{l,aa'}(p) j_{l,a'a''}(p') 
		+ y_{l,aa'a''}(p) j_{l,a''}(p'), 
			& \text{if } *=aa'a'', 
	\end{cases}
\end{equation} 
where $y_{l,a}(p)$ and $j_{l,a}(p)$ are two independent solutions to the homogeneous equation $L_{a,l}\Psi_{l,a}=0$ (here $\Psi_{l,a}$ represents one of the one-index potentials; either $\Upsilon_0$ and $\Pi_0$, depending on the value of $a$), and the other $y_{l,*}$ and $j_{l,*}$ can be calculated from these using relations given in \cite{BraamsKarney1989}. The problem of finding $N_{l,*}$ can be reduced to recursively calculating $j_{l,a}$ for $a=0,1,2$, and all $l$ of interest, however the recursive calculation is numerically non-trivial. A method for achieving accurate results is outlined in Appendix 7 of \cite{BraamsKarney1989}. Validation of the obtained $j_{l,a}$ can be done using the equation after Eq.~(A4) in \cite{BraamsKarney1989} for any $l\geq 0$, but the calculation of $y_{l,a}$ from $j_{l,a}$ requires that the recursion be performed also for $l<0$. For completeness, Table \ref{tab:y_l_table} lists some analytic expressions for $y_{l,a}$ (not contained in Eqs.~A26-A27 in \cite{BraamsKarney1989}) which are useful in validating the recursive algorithm. Note also that Eq.~(A28c) in \cite{BraamsKarney1989} has a typo, and should read $y_{1[1]2}=-(1+2z^2)\gamma/z^2$.

%%%%%%%%%%%%%%%%%%%%%%%%%%%%%%%%%%%%%%%
\begin{table}
\begin{centering}
\caption{Some analytical expressions for $y_{l,a}$ useful for validating the recursive calculation.\label{tab:y_l_table}}
\begin{tabular}{|c|c|c|c|}
\hline 
$l$\textbackslash{}$a$ & 0 & 1 & 2 \tabularnewline
\hline 
\hline 
2 & 
$-(3+2p^{2})/p^{3}$ & 
$-3\gamma/p^{3}$ & 
$-3/p^{3}$ \tabularnewline
\hline 
3 & 
$-(15\gamma/p{}^{4}+6\gamma/p^{2})$ & 
$3/p^{2}-15\gamma^{2}/p^{4}$ & 
$-(5/2)\gamma/p^{2}\times(15-15\gamma^{2}/p^{2}+21/p^{2})$  \tabularnewline
\hline 
4 & 
$-105\gamma^{2}/p{}^{5}-42\gamma^{2}/p^{3}$ & 
$-5(12\gamma/p{}^{3}+21\gamma/p{}^{5})$ & 
$-(35/2)\gamma^{2}/p^{3}\times(15-15\gamma^{2}/p^{2}$  \tabularnewline
 & $+27/p^{3}+18/p$ &  & $+21/p^{2})+15/p^{3}$  \tabularnewline
\hline 
& $-945\gamma^{3}/p{}^{6}-378\gamma^{3}/p{}^{4}$ & & 
$-(315/2)\gamma^{3}/p^{4}\times(15-15\gamma^{2}/p^{2}$ \tabularnewline
5 & $+483\gamma/p{}^{4}+258\gamma/p^{2}$ 
& $-45(7\gamma^{2}/p{}^{4} + 21\gamma^{2}/p{}^{6} +1/p^{2})$ & 
$+21/p^{2})+135\gamma/p^{4}+30\gamma/p^{2}$ \tabularnewline
 & & & $\times(15-15\gamma^{2}/p^{2}+21/p^{2})$  \tabularnewline
\hline 
\end{tabular}
\par\end{centering}
\end{table}
%%%%%%%%%%%%%%%%%%%%%%%%%%%%%%%%%%%%%%%

%%%%%%%%%%%%%%%%%%%%%%%%%%%%%%%%%%%%%%%%%%%%%%%%%%%%%%%%%%%%%%%%%%%%%%%%%%%%
%%%%%%%%%%%%%%%%%%%%%%%%%%%%%%%%%%%%%%%%%%%%%%%%%%%%%%%%%%%%%%%%%%%%%%%%%%%%
%%%%%%%%%%%%%%%%%%%%%%%%%%%%%%%%%%%%%%%%%%%%%%%%%%%%%%%%%%%%%%%%%%%%%%%%%%%%

\section*{References}
\bibliography{NORSE}

\begin{thebibliography}{47}
\expandafter\ifx\csname natexlab\endcsname\relax\def\natexlab#1{#1}\fi
\providecommand{\url}[1]{\texttt{#1}}
\providecommand{\href}[2]{#2}
\providecommand{\path}[1]{#1}
\providecommand{\DOIprefix}{doi:}
\providecommand{\ArXivprefix}{arXiv:}
\providecommand{\URLprefix}{URL: }
\providecommand{\Pubmedprefix}{pmid:}
\providecommand{\doi}[1]{\href{http://dx.doi.org/#1}{\path{#1}}}
\providecommand{\Pubmed}[1]{\href{pmid:#1}{\path{#1}}}
\providecommand{\bibinfo}[2]{#2}
\ifx\xfnm\relax \def\xfnm[#1]{\unskip,\space#1}\fi
%Type = Article
\bibitem[{Roberts and Miller(1998)}]{Roberts1998}
\bibinfo{author}{D.~A. Roberts}, \bibinfo{author}{J.~A. Miller},
  \bibinfo{journal}{Geophysical Research Letters} \bibinfo{volume}{25}
  (\bibinfo{year}{1998}) \bibinfo{pages}{607--610}.
  \DOIprefix\doi{10.1029/98GL00328}.
%Type = Article
\bibitem[{Pilipp et~al.(1987)Pilipp, Miggenrieder, Montgomery, M\"uhlh\"auser,
  Rosenbauer, and Schwenn}]{Pilipp1987a}
\bibinfo{author}{W.~G. Pilipp}, \bibinfo{author}{H.~Miggenrieder},
  \bibinfo{author}{M.~D. Montgomery}, \bibinfo{author}{K.~H. M\"uhlh\"auser},
  \bibinfo{author}{H.~Rosenbauer}, \bibinfo{author}{R.~Schwenn},
  \bibinfo{journal}{Journal of Geophysical Research: Space Physics}
  \bibinfo{volume}{92} (\bibinfo{year}{1987}) \bibinfo{pages}{1075--1092}.
  \DOIprefix\doi{10.1029/JA092iA02p01075}.
%Type = Article
\bibitem[{Van~Allen and Krimigis(1965)}]{VanAllen1965}
\bibinfo{author}{J.~A. Van~Allen}, \bibinfo{author}{S.~M. Krimigis},
  \bibinfo{journal}{Journal of Geophysical Research} \bibinfo{volume}{70}
  (\bibinfo{year}{1965}) \bibinfo{pages}{5737--5751}.
  \DOIprefix\doi{10.1029/JZ070i023p05737}.
%Type = Article
\bibitem[{Holman(1985)}]{Holman1985}
\bibinfo{author}{G.~D. Holman}, \bibinfo{journal}{Astrophysical Journal}
  \bibinfo{volume}{293} (\bibinfo{year}{1985}) \bibinfo{pages}{584--594}.
  \DOIprefix\doi{10.1086/163263}.
%Type = Article
\bibitem[{Bell et~al.(1995)Bell, Pasko, and Inan}]{Bell1995}
\bibinfo{author}{T.~F. Bell}, \bibinfo{author}{V.~P. Pasko},
  \bibinfo{author}{U.~S. Inan}, \bibinfo{journal}{Geophysical Research Letters}
  \bibinfo{volume}{22} (\bibinfo{year}{1995}) \bibinfo{pages}{2127--2130}.
  \DOIprefix\doi{10.1029/95GL02239}.
%Type = Article
\bibitem[{Gurevich et~al.(1994)Gurevich, Milikh, and
  Roussel-Dupre}]{Gurevich1994}
\bibinfo{author}{A.~Gurevich}, \bibinfo{author}{G.~Milikh},
  \bibinfo{author}{R.~Roussel-Dupre}, \bibinfo{journal}{Physics Letters A}
  \bibinfo{volume}{187} (\bibinfo{year}{1994}) \bibinfo{pages}{197 -- 203}.
  \DOIprefix\doi{10.1016/0375-9601(94)90062-0}.
%Type = Article
\bibitem[{Malka et~al.(2008)Malka, Faure, Gauduel, Lefebvre, Rousse, and
  Phuoc}]{Malka2008}
\bibinfo{author}{V.~Malka}, \bibinfo{author}{J.~Faure}, \bibinfo{author}{Y.~A.
  Gauduel}, \bibinfo{author}{E.~Lefebvre}, \bibinfo{author}{A.~Rousse},
  \bibinfo{author}{K.~T. Phuoc}, \bibinfo{journal}{Nature Physics}
  \bibinfo{volume}{4} (\bibinfo{year}{2008}) \bibinfo{pages}{447--453}.
  \DOIprefix\doi{10.1038/nphys966}.
%Type = Article
\bibitem[{Tabak et~al.(2005)Tabak, Clark, Hatchett, Key, Lasinski, Snavely,
  Wilks, Town, Stephens, Campbell, Kodama, Mima, Tanaka, Atzeni, and
  Freeman}]{Tabak2005}
\bibinfo{author}{M.~Tabak}, \bibinfo{author}{D.~S. Clark},
  \bibinfo{author}{S.~P. Hatchett}, \bibinfo{author}{M.~H. Key},
  \bibinfo{author}{B.~F. Lasinski}, \bibinfo{author}{R.~A. Snavely},
  \bibinfo{author}{S.~C. Wilks}, \bibinfo{author}{R.~P.~J. Town},
  \bibinfo{author}{R.~Stephens}, \bibinfo{author}{E.~M. Campbell},
  \bibinfo{author}{R.~Kodama}, \bibinfo{author}{K.~Mima},
  \bibinfo{author}{K.~A. Tanaka}, \bibinfo{author}{S.~Atzeni},
  \bibinfo{author}{R.~Freeman}, \bibinfo{journal}{Physics of Plasmas}
  \bibinfo{volume}{12} (\bibinfo{year}{2005}) \bibinfo{pages}{057305}.
  \DOIprefix\doi{10.1063/1.1871246}.
%Type = Article
\bibitem[{Helander et~al.(2002)Helander, Eriksson, and
  Andersson}]{Helander2002}
\bibinfo{author}{P.~Helander}, \bibinfo{author}{L.-G. Eriksson},
  \bibinfo{author}{F.~Andersson}, \bibinfo{journal}{Plasma Physics and
  Controlled Fusion} \bibinfo{volume}{44} (\bibinfo{year}{2002})
  \bibinfo{pages}{B247--62}. \DOIprefix\doi{10.1088/0741-3335/44/12B/318}.
%Type = Article
\bibitem[{Dreicer(1959)}]{Dreicer1959}
\bibinfo{author}{H.~Dreicer}, \bibinfo{journal}{Physical Review}
  \bibinfo{volume}{115} (\bibinfo{year}{1959}) \bibinfo{pages}{238--249}.
  \DOIprefix\doi{10.1103/PhysRev.115.238}.
%Type = Article
\bibitem[{Dreicer(1960)}]{Dreicer1960}
\bibinfo{author}{H.~Dreicer}, \bibinfo{journal}{Physical Review}
  \bibinfo{volume}{117} (\bibinfo{year}{1960}) \bibinfo{pages}{329--342}.
  \DOIprefix\doi{10.1103/PhysRev.117.329}.
%Type = Article
\bibitem[{Rosenbluth and Putvinski(1997)}]{RosenbluthPutvinski1997}
\bibinfo{author}{M.~Rosenbluth}, \bibinfo{author}{S.~Putvinski},
  \bibinfo{journal}{Nuclear Fusion} \bibinfo{volume}{37} (\bibinfo{year}{1997})
  \bibinfo{pages}{1355--1362}. \DOIprefix\doi{10.1088/0029-5515/37/10/I03}.
%Type = Article
\bibitem[{Boozer(2015)}]{Boozer2015}
\bibinfo{author}{A.~H. Boozer}, \bibinfo{journal}{Physics of Plasmas}
  \bibinfo{volume}{22} (\bibinfo{year}{2015}) \bibinfo{pages}{032504}.
  \DOIprefix\doi{10.1063/1.4913582}.
%Type = Article
\bibitem[{Hirvijoki et~al.(2015)Hirvijoki, Pusztai, Decker, Embr\'eus, Stahl,
  and F\"ul\"op}]{HirvijokiBump2015}
\bibinfo{author}{E.~Hirvijoki}, \bibinfo{author}{I.~Pusztai},
  \bibinfo{author}{J.~Decker}, \bibinfo{author}{O.~Embr\'eus},
  \bibinfo{author}{A.~Stahl}, \bibinfo{author}{T.~F\"ul\"op},
  \bibinfo{journal}{Journal of Plasma Physics} \bibinfo{volume}{81}
  (\bibinfo{year}{2015}) \bibinfo{pages}{475810502}.
  \DOIprefix\doi{10.1017/S0022377815000513/}.
%Type = Article
\bibitem[{Decker et~al.(2016)Decker, Hirvijoki, Embreus, Peysson, Stahl,
  Pusztai, and F\"ul\"op}]{DeckerBump2016}
\bibinfo{author}{J.~Decker}, \bibinfo{author}{E.~Hirvijoki},
  \bibinfo{author}{O.~Embreus}, \bibinfo{author}{Y.~Peysson},
  \bibinfo{author}{A.~Stahl}, \bibinfo{author}{I.~Pusztai},
  \bibinfo{author}{T.~F\"ul\"op}, \bibinfo{journal}{Plasma Physics and
  Controlled Fusion} \bibinfo{volume}{58} (\bibinfo{year}{2016})
  \bibinfo{pages}{025016}. \DOIprefix\doi{10.1088/0741-3335/58/2/025016}.
%Type = Article
\bibitem[{Pokol et~al.(2008)Pokol, F\"ul\"op, and Lisak}]{Pokol2008}
\bibinfo{author}{G.~Pokol}, \bibinfo{author}{T.~F\"ul\"op},
  \bibinfo{author}{M.~Lisak}, \bibinfo{journal}{Plasma Physics and Controlled
  Fusion} \bibinfo{volume}{50} (\bibinfo{year}{2008}) \bibinfo{pages}{045003}.
  \DOIprefix\doi{10.1088/0741-3335/50/4/045003}.
%Type = Article
\bibitem[{Pokol et~al.(2014)Pokol, K\'om\'ar, Budai, Stahl, and
  F\"ul\"op}]{Pokol2014}
\bibinfo{author}{G.~I. Pokol}, \bibinfo{author}{A.~K\'om\'ar},
  \bibinfo{author}{A.~Budai}, \bibinfo{author}{A.~Stahl},
  \bibinfo{author}{T.~F\"ul\"op}, \bibinfo{journal}{Physics of Plasmas}
  \bibinfo{volume}{21} (\bibinfo{year}{2014}) \bibinfo{pages}{102503}.
  \DOIprefix\doi{10.1063/1.4895513}.
%Type = Article
\bibitem[{Chiu et~al.(1998)Chiu, Rosenbluth, Harvey, and Chan}]{Chiu1998}
\bibinfo{author}{S.~Chiu}, \bibinfo{author}{M.~Rosenbluth},
  \bibinfo{author}{R.~Harvey}, \bibinfo{author}{V.~Chan},
  \bibinfo{journal}{Nuclear Fusion} \bibinfo{volume}{38} (\bibinfo{year}{1998})
  \bibinfo{pages}{1711--1721}. \DOIprefix\doi{10.1088/0029-5515/38/11/309}.
%Type = Techreport
\bibitem[{Decker and Peysson(2004)}]{Decker2004}
\bibinfo{author}{J.~Decker}, \bibinfo{author}{Y.~Peysson},
  \bibinfo{title}{{DKE}: {A} fast numerical solver for the {3D} drift kinetic
  equation}, \bibinfo{type}{Technical Report}
  \bibinfo{number}{EUR-CEA-FC-1736}, Euratom-CEA, \bibinfo{year}{2004}.
%Type = Article
\bibitem[{Landreman et~al.(2014)Landreman, Stahl, and
  F\"ul\"op}]{CODEPaper2014}
\bibinfo{author}{M.~Landreman}, \bibinfo{author}{A.~Stahl},
  \bibinfo{author}{T.~F\"ul\"op}, \bibinfo{journal}{Computer Physics
  Communications} \bibinfo{volume}{185} (\bibinfo{year}{2014})
  \bibinfo{pages}{847 -- 855}. \DOIprefix\doi{10.1016/j.cpc.2013.12.004}.
%Type = Article
\bibitem[{McCoy et~al.(1981)McCoy, Mirin, and Killeen}]{McCoy1981}
\bibinfo{author}{M.~McCoy}, \bibinfo{author}{A.~Mirin},
  \bibinfo{author}{J.~Killeen}, \bibinfo{journal}{Computer Physics
  Communications} \bibinfo{volume}{24} (\bibinfo{year}{1981})
  \bibinfo{pages}{37 -- 61}. \DOIprefix\doi{10.1016/0010-4655(81)90105-3}.
%Type = Article
\bibitem[{Chac\'on et~al.(2000)Chac\'on, Barnes, Knoll, and Miley}]{Chacon2000}
\bibinfo{author}{L.~Chac\'on}, \bibinfo{author}{D.~Barnes},
  \bibinfo{author}{D.~Knoll}, \bibinfo{author}{G.~Miley},
  \bibinfo{journal}{Journal of Computational Physics} \bibinfo{volume}{157}
  (\bibinfo{year}{2000}) \bibinfo{pages}{654 -- 682}.
  \DOIprefix\doi{10.1006/jcph.1999.6395}.
%Type = Article
\bibitem[{Pareschi et~al.(2000)Pareschi, Russo, and Toscani}]{Pareschi2000}
\bibinfo{author}{L.~Pareschi}, \bibinfo{author}{G.~Russo},
  \bibinfo{author}{G.~Toscani}, \bibinfo{journal}{Journal of Computational
  Physics} \bibinfo{volume}{165} (\bibinfo{year}{2000}) \bibinfo{pages}{216 --
  236}. \DOIprefix\doi{10.1006/jcph.2000.6612}.
%Type = Article
\bibitem[{Weng et~al.(2006)Weng, Sheng, He, Wu, Dong, and Zhang}]{Weng2006}
\bibinfo{author}{S.-M. Weng}, \bibinfo{author}{Z.-M. Sheng},
  \bibinfo{author}{M.-Q. He}, \bibinfo{author}{H.-C. Wu},
  \bibinfo{author}{Q.-L. Dong}, \bibinfo{author}{J.~Zhang},
  \bibinfo{journal}{Physics of Plasmas} \bibinfo{volume}{13}
  (\bibinfo{year}{2006}) \bibinfo{pages}{113302}.
  \DOIprefix\doi{10.1063/1.2370725}.
%Type = Article
\bibitem[{Pataki and Greengard(2011)}]{Pataki2011}
\bibinfo{author}{A.~Pataki}, \bibinfo{author}{L.~Greengard},
  \bibinfo{journal}{Journal of Computational Physics} \bibinfo{volume}{230}
  (\bibinfo{year}{2011}) \bibinfo{pages}{7840 -- 7852}.
  \DOIprefix\doi{10.1016/j.jcp.2011.07.005}.
%Type = Article
\bibitem[{Yoon and Chang(2014)}]{Yoon2014}
\bibinfo{author}{E.~S. Yoon}, \bibinfo{author}{C.~S. Chang},
  \bibinfo{journal}{Physics of Plasmas} \bibinfo{volume}{21}
  (\bibinfo{year}{2014}) \bibinfo{pages}{032503}.
  \DOIprefix\doi{10.1063/1.4867359}.
%Type = Article
\bibitem[{Taitano et~al.(2015)Taitano, Chac\'on, Simakov, and
  Molvig}]{Taitano2015}
\bibinfo{author}{W.~Taitano}, \bibinfo{author}{L.~Chac\'on},
  \bibinfo{author}{A.~Simakov}, \bibinfo{author}{K.~Molvig},
  \bibinfo{journal}{Journal of Computational Physics} \bibinfo{volume}{297}
  (\bibinfo{year}{2015}) \bibinfo{pages}{357 -- 380}.
  \DOIprefix\doi{10.1016/j.jcp.2015.05.025}.
%Type = Article
\bibitem[{Hirvijoki et~al.(2015)Hirvijoki, Candy, Belli, and
  Embreus}]{HirvijokiRBF2015}
\bibinfo{author}{E.~Hirvijoki}, \bibinfo{author}{J.~Candy},
  \bibinfo{author}{E.~Belli}, \bibinfo{author}{O.~Embreus},
  \bibinfo{journal}{Physics Letters A} \bibinfo{volume}{379}
  (\bibinfo{year}{2015}) \bibinfo{pages}{2735--2739}.
  \DOIprefix\doi{10.1016/j.physleta.2015.08.010}.
%Type = Article
\bibitem[{Nuga and Fukuyama(2011)}]{Nuga2011}
\bibinfo{author}{H.~Nuga}, \bibinfo{author}{A.~Fukuyama},
  \bibinfo{journal}{Progress in Nuclear Science and Technology}
  \bibinfo{volume}{2} (\bibinfo{year}{2011}) \bibinfo{pages}{78--84}.
  \DOIprefix\doi{10.15669/pnst.2.78}.
%Type = Unpublished
\bibitem[{Petrov and Harvey(2009)}]{Petrov2009}
\bibinfo{author}{Y.~Petrov}, \bibinfo{author}{R.~W. Harvey},
  \bibinfo{title}{Benchmarking the fully relativistic collision operator in
  {CQL3D}}, \bibinfo{year}{2009}. \URLprefix
  \url{http://www.compxco.com/CompX-2009-1_Fully-Rel.pdf},
  \bibinfo{note}{{CompX} report {CompX}-2009-1}.
%Type = Article
\bibitem[{Connor and Hastie(1975)}]{ConnorHastie1975}
\bibinfo{author}{J.~Connor}, \bibinfo{author}{R.~Hastie},
  \bibinfo{journal}{Nuclear Fusion} \bibinfo{volume}{15} (\bibinfo{year}{1975})
  \bibinfo{pages}{415--424}. \DOIprefix\doi{10.1088/0029-5515/15/3/007}.
%Type = Article
\bibitem[{Beliaev and Budker(1956)}]{BeliaevBudker1956}
\bibinfo{author}{S.~T. Beliaev}, \bibinfo{author}{G.~I. Budker},
  \bibinfo{journal}{Soviet Physics-Doklady} \bibinfo{volume}{1}
  (\bibinfo{year}{1956}) \bibinfo{pages}{218}.
%Type = Article
\bibitem[{Braams and Karney(1987)}]{BraamsKarney1987}
\bibinfo{author}{B.~J. Braams}, \bibinfo{author}{C.~F.~F. Karney},
  \bibinfo{journal}{Physical Review Letters} \bibinfo{volume}{59}
  (\bibinfo{year}{1987}) \bibinfo{pages}{1817--1820}.
  \DOIprefix\doi{10.1103/PhysRevLett.59.1817}.
%Type = Article
\bibitem[{Braams and Karney(1989)}]{BraamsKarney1989}
\bibinfo{author}{B.~J. Braams}, \bibinfo{author}{C.~F.~F. Karney},
  \bibinfo{journal}{Physics of Fluids B: Plasma Physics} \bibinfo{volume}{1}
  (\bibinfo{year}{1989}) \bibinfo{pages}{1355--1368}.
  \DOIprefix\doi{10.1063/1.858966}.
%Type = Article
\bibitem[{Andersson et~al.(2001)Andersson, Helander, and
  Eriksson}]{Andersson2001}
\bibinfo{author}{F.~Andersson}, \bibinfo{author}{P.~Helander},
  \bibinfo{author}{L.-G. Eriksson}, \bibinfo{journal}{Physics of Plasmas}
  \bibinfo{volume}{8} (\bibinfo{year}{2001}) \bibinfo{pages}{5221--5229}.
  \DOIprefix\doi{10.1063/1.1418242}.
%Type = Book
\bibitem[{Helander and Sigmar(2002)}]{HelanderSigmar2002}
\bibinfo{author}{P.~Helander}, \bibinfo{author}{D.~J. Sigmar},
  \bibinfo{title}{Collisional Transport in Magnetized Plasmas},
  \bibinfo{publisher}{Cambridge University Press}, \bibinfo{year}{2002}.
%Type = Article
\bibitem[{Rosenbluth et~al.(1957)Rosenbluth, MacDonald, and
  Judd}]{Rosenbluth1957}
\bibinfo{author}{M.~N. Rosenbluth}, \bibinfo{author}{W.~M. MacDonald},
  \bibinfo{author}{D.~L. Judd}, \bibinfo{journal}{Physical Review}
  \bibinfo{volume}{107} (\bibinfo{year}{1957}) \bibinfo{pages}{1--6}.
  \DOIprefix\doi{10.1103/PhysRev.107.1}.
%Type = Article
\bibitem[{Saad and Schultz(1986)}]{SaadSchultz1986}
\bibinfo{author}{Y.~Saad}, \bibinfo{author}{M.~H. Schultz},
  \bibinfo{journal}{{SIAM} Journal on Scientific and Statistical Computing}
  \bibinfo{volume}{7} (\bibinfo{year}{1986}) \bibinfo{pages}{856--869}.
  \DOIprefix\doi{10.1137/0907058}.
%Type = Article
\bibitem[{Weng et~al.(2008)Weng, Sheng, He, Zhang, Norreys, Sherlock, and
  Robinson}]{Weng2008}
\bibinfo{author}{S.~M. Weng}, \bibinfo{author}{Z.~M. Sheng},
  \bibinfo{author}{M.~Q. He}, \bibinfo{author}{J.~Zhang},
  \bibinfo{author}{P.~A. Norreys}, \bibinfo{author}{M.~Sherlock},
  \bibinfo{author}{A.~P.~L. Robinson}, \bibinfo{journal}{Physical Review
  Letters} \bibinfo{volume}{100} (\bibinfo{year}{2008})
  \bibinfo{pages}{185001}. \DOIprefix\doi{10.1103/PhysRevLett.100.185001}.
%Type = Misc
\bibitem[{Weng(2016)}]{Weng2016}
\bibinfo{author}{S.~M. Weng}, \bibinfo{title}{private communication},
  \bibinfo{year}{2016}.
%Type = Article
\bibitem[{Stahl et~al.(2016)Stahl, Embr\'eus, Papp, Landreman, and
  F\"ul\"op}]{Stahl2016}
\bibinfo{author}{A.~Stahl}, \bibinfo{author}{O.~Embr\'eus},
  \bibinfo{author}{G.~Papp}, \bibinfo{author}{M.~Landreman},
  \bibinfo{author}{T.~F\"ul\"op}, \bibinfo{journal}{Nuclear Fusion}
  \bibinfo{volume}{56} (\bibinfo{year}{2016}) \bibinfo{pages}{112009}.
  \DOIprefix\doi{10.1088/0029-5515/56/11/112009}.
%Type = Article
\bibitem[{Smith et~al.(2005)Smith, Helander, Eriksson, and
  F\"ul\"op}]{Smith2005}
\bibinfo{author}{H.~Smith}, \bibinfo{author}{P.~Helander},
  \bibinfo{author}{L.-G. Eriksson}, \bibinfo{author}{T.~F\"ul\"op},
  \bibinfo{journal}{Physics of Plasmas} \bibinfo{volume}{12}
  (\bibinfo{year}{2005}) \bibinfo{pages}{122505}.
  \DOIprefix\doi{10.1063/1.2148966}.
%Type = Article
\bibitem[{Mart\'in-Sol\'is et~al.(1998)Mart\'in-Sol\'is, Alvarez, S\'anchez,
  and Esposito}]{MartinSolis1998}
\bibinfo{author}{J.~R. Mart\'in-Sol\'is}, \bibinfo{author}{J.~D. Alvarez},
  \bibinfo{author}{R.~S\'anchez}, \bibinfo{author}{B.~Esposito},
  \bibinfo{journal}{Physics of Plasmas} \bibinfo{volume}{5}
  (\bibinfo{year}{1998}) \bibinfo{pages}{2370--2377}.
  \DOIprefix\doi{10.1063/1.872911}.
%Type = Article
\bibitem[{Stahl et~al.(2015)Stahl, Hirvijoki, Decker, Embr\'eus, and
  F\"{u}l\"{o}p}]{Stahl2015}
\bibinfo{author}{A.~Stahl}, \bibinfo{author}{E.~Hirvijoki},
  \bibinfo{author}{J.~Decker}, \bibinfo{author}{O.~Embr\'eus},
  \bibinfo{author}{T.~F\"{u}l\"{o}p}, \bibinfo{journal}{Physical Review
  Letters} \bibinfo{volume}{114} (\bibinfo{year}{2015})
  \bibinfo{pages}{115002}. \DOIprefix\doi{10.1103/PhysRevLett.114.115002}.
%Type = Article
\bibitem[{Aleynikov and Breizman(2015)}]{Aleynikov2015}
\bibinfo{author}{P.~Aleynikov}, \bibinfo{author}{B.~N. Breizman},
  \bibinfo{journal}{Physical Review Letters} \bibinfo{volume}{114}
  (\bibinfo{year}{2015}) \bibinfo{pages}{155001}.
  \DOIprefix\doi{10.1103/PhysRevLett.114.155001}.
%Type = Article
\bibitem[{Liu et~al.(2016)Liu, Brennan, Bhattacharjee, and Boozer}]{Liu2016}
\bibinfo{author}{C.~Liu}, \bibinfo{author}{D.~P. Brennan},
  \bibinfo{author}{A.~Bhattacharjee}, \bibinfo{author}{A.~H. Boozer},
  \bibinfo{journal}{Physics of Plasmas} \bibinfo{volume}{23}
  (\bibinfo{year}{2016}) \bibinfo{pages}{010702}.
  \DOIprefix\doi{10.1063/1.4938510}.
%Type = Article
\bibitem[{Coppi et~al.(1976)Coppi, Pegoraro, Pozzoli, and Rewoldt}]{Coppi1976}
\bibinfo{author}{B.~Coppi}, \bibinfo{author}{F.~Pegoraro},
  \bibinfo{author}{R.~Pozzoli}, \bibinfo{author}{G.~Rewoldt},
  \bibinfo{journal}{Nuclear Fusion} \bibinfo{volume}{16} (\bibinfo{year}{1976})
  \bibinfo{pages}{309}. \URLprefix
  \url{http://stacks.iop.org/0029-5515/16/i=2/a=014}.

\end{thebibliography}

%%%%%%%%%%%%%%%%%%%%%%%%%%%%%%%%%%%%%%%%%%%%%%%%%%%%%%%%%%%%%%%%%%%%%%%%%%%%

\end{document}